\documentclass[]{interact}
\usepackage{epstopdf}
\usepackage[caption=false]{subfig}

\usepackage{natbib}
\theoremstyle{plain}

\theoremstyle{definition}

\theoremstyle{remark}

\usepackage{amsmath}
\usepackage{amssymb}
\usepackage{booktabs}
\usepackage{array}
\usepackage{subcaption}
\usepackage{enumitem}
\usepackage{microtype}

\usepackage{array}
\usepackage{threeparttable}
\usepackage{multirow}
\usepackage{xcolor}
\usepackage{rotating}
\usepackage{makecell}
\usepackage{caption}
\usepackage{subcaption}
\usepackage{comment}

\newcolumntype{L}[1]{>{\raggedright\arraybackslash}p{#1}}
\newcolumntype{C}[1]{>{\centering\arraybackslash}p{#1}}

\usepackage[usegeometry]{typearea}
\newenvironment{uselscape}{
\KOMAoptions{paper=landscape, DIV=current}
\newgeometry{top=3 cm, left = 3.2 cm, right = 3.2 cm, bottom = 3 cm}
}{\clearpage}

\begin{document}

\articletype{}

\title{A Hybrid Discrete-Event and Agent-Based Simulation Approach to Model Circular Supply Chains in Healthcare: A Case Study of Laparoscopic Scissors }

\author{
\name{Mohd Shoaib\textsuperscript{a,b}\thanks{CONTACT Mohd Shoaib. Email: mohd.shoaib@ntu.ac.uk}, Antuela Tako\textsuperscript{a}, and Shahin Rahimifard\textsuperscript{c}}
\affil{\textsuperscript{a}Nottingham Business School, Nottingham Trent University, Nottingham, UK; \textsuperscript{b}The Bartlett School of Environment, Energy and Resources, Energy Institute, University College London, U.K.; \textsuperscript{c}Centre for Sustainable Manufacturing and Recycling Technologies,
Wolfson School of Mechanical Electrical and Manufacturing Engineering, Loughborough University, U.K.
}}

\maketitle

\begin{abstract}
Circular healthcare supply chains are inherently complex,  characterised by interdependencies among their actors and high uncertainty in product flows and performance.  Current methods used to predict the outcomes of transitioning to circular economy (CE) are limited and mostly static. This paper demonstrates the use of simulation  to assess the effect of introducing circular products and the implications across the healthcare supply chain accounting for variability. The laparoscopic scissors supply chain is chosen as a case study example.  
To the best of our knowledge, this is the first study that assesses the implications of introducing circular product (medical devices) designs at both the individual supply chain member and overall system level. The model can be also used to inform optimal inventory strategies for hospitals,  to ensure that patient safety and hospital operations are maintained. Our findings suggest that adopting circular products can reduce the environmental impact, but to achieve significant reductions in both cost and emissions, it requires significant upfront investment. 
We discuss the theoretical and practical implications of our study in developing tools to support the transition to CE. 
\end{abstract}

\section*{Practitioner Summary}

Adopting circular economy strategies in healthcare supply chains is complex because the impacts of these strategies on costs, emissions, and operational performance are unknown and subjected to  uncertainty. Decision-makers, therefore, need tools that help them understand the long-term implications of transitioning from linear to circular systems under uncertainty.

This study presents a hybrid Agent-based and Discrete Event simulation model that supports decision-making by evaluating the impacts of different circular strategies for medical devices. Using laparoscopic scissors as a case study, the model evaluates the costs and carbon emissions of three circular strategies — remanufactured, hybrid, and fully reusable laparoscopic scissors — while capturing the operational complexity and variability introduced by product returns and reprocessing activities.

The results show that no single approach dominates across the scenarios tested. Further, the model identifies the environmental and economic tipping points for each strategy compared with the current linear system. The study provides healthcare managers with actionable operational guidance, specifying, for example, the sterilisation capacity and staffing levels required to handle specific throughput demands for reusable instruments. Similarly, for policymakers, the study highlights the need for policy instruments that subsidise the high upfront procurement costs associated with hybrid instruments.

\begin{keywords}
Simulation; circular economy; supply chain; Discrete event simulation; small medical devices; decision support; inventory management
\end{keywords}

\section{Introduction}
\label{sec:intro}

The transition to a circular economy (CE) demands changes across product design, manufacturing, end-of-life (EoL) management, supply chain (SC) configuration, and business models. These changes extend beyond individual products to encompass supply chains, business models, and operational processes, with implications for economic performance, environmental impact, and system efficiency \citep{Bimpizas-Pinis2022, Linder2017, Fernndez2025}.  Consequently, decision-making tools are needed to assess how circular strategies affect system performance under uncertainty and over time, to  inform their adoption.

A range of approaches has been used to study CE transitions, including life cycle-based assessments and optimisation-based modelling. Life cycle assessment (LCA) quantifies the environmental impacts of alternative product systems and is frequently paired with life cycle cost analysis (LCCA) to compare costs, particularly in healthcare, where studies have compared single-use (SU) and reusable (RU) or hybrid medical devices (e.g. \cite{rizan2022environmental}). Optimisation models support decisions on supply chain design, recovery processes, and material flows \cite{Konstantaras2021, YahyapourGanji2025}. While these approaches provide valuable insights, they are less suitable for system-wide evaluation.

Furthermore, the adoption of circular products leads to the structural reconfiguration of supply chains, including the introduction of new actors (e.g. repairers and remanufacturers) and reverse flows such as return and reprocessing loops. These changes introduce additional sources of operational variability—such as lead time uncertainty and resource constraints—that are absent in linear systems. This is particularly important in healthcare delivery, where instrument unavailability may determine whether a surgical procedure can proceed, with direct consequences for service delivery and patient risk.

Capturing these dynamics requires a modelling approach that can explicitly represent the flow of devices  across the supply chain over time accounting for the stochastic patterns in stock availability and replenishment times. 
Simulation-based approaches provide such capabilities and are therefore well suited to analyse CE transitions in complex operational systems. Methods such as discrete-event simulation (DES), agent-based simulation (ABS), and hybrid approaches allow for the explicit modelling of process flows, interactions between system components, and stochastic variability over time \citep{Brailsford2009, Tako2012}. Moreover, simulation is widely recognised as an effective approach for modelling supply chains \citep{Chilmon2020}, offering a risk-free environment to conduct what-if analyses of future or non-existent scenarios, as is the case with the transitioning to circular economy. These capabilities make simulation a highly suitable and flexible method for use in assessing CE-driven transformations in small medical device (SMD) supply chains \cite{Moon2017, Sassanelli2020}.

Reflecting this, simulation has been applied in multiple domains to analyse CE transitions \cite{Amico2024, Lieder2017, Charnley2019}, but its use in healthcare remains limited to testing business model changes \cite{Guzzo2019} or specific EoL strategies \cite{shoaibesc}. Existing healthcare simulation studies remain limited in three respects. First, they lack network-level supply chain representations capturing member-level heterogeneity and operational uncertainty. Second, they do not consider alternative circular product designs. Third, they do not model reuse as a multi-stage operational process but treat it as either an adoption decision \cite{Walzberg2021} or as the reintroduction of products following EoL treatment \cite{Lieder2017}. This limitation is particularly significant in healthcare, where reusing instruments such as laparoscopic scissors requires strict decontamination processes after each use. These resource-intensive steps consume time, introduce operational variability, and directly impact both cost and environmental performance

To address these gaps, this study develops a hybrid simulation model (a DES that utilises agent-based simulation features) of a four-echelon SMD supply and recovery chain, using laparoscopic scissors as a case study. The model adopts a whole-systems perspective, representing supply chain members as autonomous agents and capturing their interactions alongside the internal operational processes including production, distribution, use, and multi-stage reprocessing, that CE led transition introduces. In particular, detailed operational flows including disassembly, cleaning, decontamination, and reassembly are incorporated into the model to enable the dynamic analysis of the operational implications of circular strategies under uncertainty.

The study evaluates the performance of linear and circular supply chain configurations representing different product designs across multiple scenarios, examining trade-offs between  cost and environmental implications to identify conditions under which circular strategies are feasible. Furthermore, operational performance measures are assessed using a simulation–optimisation approach to support hospital inventory planning under uncertainty. Sensitivity analysis is conducted to identify key parameters influencing the feasibility of circular strategies.

The contributions of this paper are threefold. First, we develop a novel hybrid simulation framework,  combining discrete-event and agent-based simulation, that operationalises reuse as a dynamic, multi-stage operational flow rather than a static product life extension. Second, this is the first study, in our knowledge, to evaluate CE transition strategies in healthcare supply chains of small medical devices at both system- and member-level for  different circular product designs. Third, the study quantifies resource and inventory implications when adopting circular product designs to support managerial decisions, identify the pricing levels when circular products become viable, and inform government subsidies.

The remainder of the paper is organised as follows. Section \ref{sec:lit-review}, considers the main approaches used to model the transition to circular economy. Next, Section \ref{sec:methodology}, provides a detailed description of the modelling approach adopted and model architecture Sections \ref{sec:results} and 5,  present the results of simulation experiments carried out to assess the implications of transitioning to CE in healthcare SCs. Finally, Section \ref{sec:conclusion}, concludes the article. 

\section{Related Work}
\label{sec:lit-review}

Research from multiple sectors has shown that the adoption of CE principles can yield overall positive environmental, operational, and economic benefits \cite{Kramer2023, Lewandowski2016}. However, adoption of CE demands product design changes, supply chain network restructuring, operational flow adjustments, business model shifts, and new EoL management strategies among others, which could result in implications for the supply network. Thus, understanding these implications before real-world adoption is crucial to avoid unintended consequences and requires approaches that can capture key performance measures across different scenarios.

\subsection{Methods Used to Measure Implications of Adopting Circular Economy Principles}
\label{sec:lit-review/other-methodologies}

Various methodologies have been used to measure the implications of adopting CE principles including such as LCAs and operational research methods such as mathematical modelling, and simulation \cite{Sassanelli2019}. These methods differ in scope, detail, outcomes, and ability to represent the systems complexity in dynamic operational settings.

\subsubsection{Life Cycle Assessment (LCA)}

LCA is one of the most widely employed analytical methodologies to measure the environmental impact of transitioning to CE \cite{Sassanelli2019, Walzberg2020}. LCA studies generally include all the processes associated with a product's lifecycle from raw material extraction and production to EoL disposal \cite{Guinee2017, Curran2013}. An inventory of material and energy inputs and outputs is compiled and then translated into environmental impacts, such as global warming potential, acidification potential, eutrophication potential, and ozone depletion \cite{Drew2022}. Although LCAs are widely used for environmental assessments across sectors, we restrict the focus of this review to the healthcare sector. We direct interested readers to \cite{Haupt2017, Pea2021} for more information on the topic.

In healthcare, LCAs have been undertaken to assess the implications of introducing CE principles from both a process-based and product-perspective. The former are concerned with assessments of surgical procedures i.e. cardiac surgery, open and laparoscopic; management of type-2 diabetes;laboratory testing and imaging \cite{Drew2022}. The latter involve assessments of healthcare products whereby comparison of SU against RU and/or hybrid medical devices. For example, 
Lieden et al. \cite{Leiden2020} compare RU spinal fusion surgical instruments with SU options and found that the SU option was more environmentally-friendly compared to reusable instruments, highlighting that cleaning and sterilisation processes are high carbon emission hotspots. Similar results have been reported for laparoscopic instruments used for cholecystectomies \cite{Boberg2022, rizan2022environmental}. 
Boberg et al. \cite{Boberg2022} found that reusable devices, i.e. trocar systems used in laparoscopic cholecystectomies, were financially and environmentally better compared to SU ones, while Rizan and Bhutta \cite{rizan2022environmental} found that hybrid instruments outperformed SU instruments considering both cost and carbon emissions.

\subsubsection{Other Methodologies}

Systems thinking has been used to examine structural and behavioural conditions that influence circular transitions, providing insights into system interdependence and feedback mechanisms that shape the transition to circularity \cite{Bassi2021, Crdoba-Pachn2025, Gajanayake2025}. While useful for understanding system structure and policy dynamics at a high level, systems thinking approaches do not quantify operational performance or supply chain behaviour.

Mathematical modelling and optimisation provides decision support for specific CE problems, including reverse logistics and closed-loop supply chain design. Suzanne et al. \cite{Suzanne2020} present a conceptual optimisation framework for circular supply chain network design, highlighting that integrated decisions on product design, recovery processes, and supply chain configuration are required for transitioning to CE. Konstantaras et al. \cite{Konstantaras2021} developed a linear programming model to optimise the total cost of a circular supply chain integrating manufacturing, remanufacturing, and repair activities. In the healthcare sector specifically, Yahyapour Ganji et al. \cite{YahyapourGanji2025} developed and solved a multi-stage data-driven optimisation model for designing a circular, resilient, and responsive supply chain for haemodialysis devices. A broader discussion of optimisation decisions for CE transitions and additional methodological approaches, including material flow analysis, input-output modelling, and multi-criteria decision analysis, is beyond the scope of this work; interested readers may refer to \cite{Sassanelli2019, Echefaj2025}.

While these methodologies offer powerful approaches for CE transition assessments, they are limited in their capacity to capture both structural system dynamics and stochastic operational behaviors simultaneously. Systems thinking is useful for understanding system interdependence and structural barriers but does not quantify operational performance or supply chain behaviour. Mathematical modelling and optimisation-based approaches support decision-making but have limited ability to capture randomness and handle the stochastic and dynamic characteristics associated with CE transitions.

\subsubsection{Simulation Modelling}

Simulation modelling is a well-recognised approach to analyse and optimise the performance of complex systems with inherent uncertainty and dynamic behaviour, such as health systems and SCs \cite{Brailsford2009, Chilmon2020}. It is particularly suitable for studying the dynamics of systems that do not yet exist \cite{Carson2005}. It allows a risk-free environment for exploring system behaviour under different scenarios, offering valuable insights for decision-making and system optimisation prior to real-world implementation \cite{Moon2017}. In the field of operational research, the primary simulation methodologies include SD, DES, and ABS \cite{Katsaliaki2011}. In the literature these methodologies have been widely used as decision support tools across almost all disciplines \cite{Taylor2009}. Moreover, sometimes these methods are integrated or combined to draw upon the strengths of the combined methodologies \cite{Brailsford2019} when a single method is not sufficient. 

In the context of CE, simulation-based research focusses on evaluating i) impacts of different EoL management strategies, ii) the effect of business model change, and iii) explore the combination of both EoL strategies and  business model change. These CE strategies are assessed using common performance indicators that include carbon dioxide emissions, flow of products to landfills (for disposal),  new product utilisation, and the effects on lead times among others.

Amico et al. \cite{Amico2024} developed a DES model of the supply chain for global apparel polyester jackets and compare the effects of adopting different EoL strategies (reuse, remanufacture, and recycle) on environmental and economic outcomes. They find that reuse is a more profitable strategy; however, reuse and recycle strategies jointly result in better economic and environmental performance at system-level. 
Huster et al. \cite{Huster2022} developed a DES model to understand the impact of remanufacturing of used electric vehicle batteries (EVBs) on demand for new EVBs, concluding that remanufacturing could potentially reduce new battery demand. In another study, Charnley et al. \cite{Charnley2019} used DES to develop a decision support system for a hypothetical EVB remanufacturing facility, finding that variation in the quality of returned units significantly affects process times and throughput.

 Guzzo et al. \cite{Guzzo2019} tested a sharing platform (CE strategy) for low-utilisation consumable and durable medical products in a hospital using SD simulation. The authors showed that the sharing strategy improved system performance with a clear reduction in both new product requirement and unmet demand compared with a linear scenario. In the fashion industry, de Olan˜eta et al. \cite{deOlaeta2023} developed a DES to optimise inventory and cash flows of a startup operating on a subscription-fee based model, establishing that the performance of this business model is sensitive to inventory policy and customer subscription dynamics. Finally, Lieder et al. \cite{Lieder2017} analysed life-cycle costs and carbon emissions for different EoL strategies, including reuse, remanufacturing, and recycling options — and business models such as buy-back, leasing, and pay-per-use schemes, using a hybrid ABS-DES model with a washing machine as the case study product. Their analysis demonstrated that the choice of business model has a significant impact on environmental and economic performance measures.

\subsection{Gaps from the Existing Literature for Assessing Circular Economy Transition in Healthcare}
From the studies discussed, it can be concluded that modelling approaches for assessing CE in the context of medical devices are limited. Current studies focus primarily on testing the feasibility of transitioning to a CE set up. The analysis is primarily deterministic using methods such as LCAs and mathematical modelling. Few studies use simulation in the healthcare sector, and particularly for medical devices, but these focus primarily on internal hospital operations, patient flow, or localized inventory management. We found no studies that consider the operational implications of CE implementation on the supply and recovery chain, and more so in the context of medical devices. Therefore, our paper provides a novel study that conceptualises the CE supply chain of medical devices in two parts the supply and recovery chain. Moreover, we argue that using a simulation approach it is possible to represent dynamically the interaction of parts of the system over time and in doing so it enables the assessment of operational implications alongside carbon emissions and cost implications in the healthcare supply chains.

A summary of the literature discussed in this section is provided in Table \ref{tab:litreview} in Appendix \ref{app:lit-summary}.

We next introduce the methodological approach in developing the simulation model that is specifically designed to represent the supply and recovery chain of medical devices.

\section{Methodology}
\label{sec:methodology}

This section describes the study context, modelling approach, system representation, and simulation design used to evaluate CE transitions in SMD supply chains.

\subsection{Study context and scope}

Laparoscopic scissors were selected as the case study product given their widespread use in NHS surgical procedures and their relevance to NHS sustainability targets.
Laparoscopic scissors used within the UK NHS are predominantly SU and operate within the LE-based system. Thus, after each procedure these are disposed of as clinical waste through high-temperature incineration, contributing to both emissions and material waste.
To mitigate these impacts, the adoption of circular design options such as fully reusable, hybrid (reusable handle with a disposable blade), and remanufactured laparoscopic scissors can be considered.

However, replacing SU laparoscopic scissors with circular alternatives changes the supply chain configuration by introducing reverse logistics and reprocessing loops. To capture these dynamics, we conceptualised the end-to-end operational flows for the base case linear economy and proposed three circular supply chain strategies. These conceptual models are illustrated in Figure~\ref{fig:conceptual} and detailed below.

\subsubsection{Linear Economy: single-use Laparoscopic Scissors}
 
The LE scenario represents the current practice of using SU laparoscopic
scissors that are disposed of after a single use cycle
(Figure~\ref{fig:conceptual}a).
Scissors are manufactured and shipped for distribution to the distribution centre (DC) and subsequently to hospitals via truck transport.
After use at hospitals, instruments enter a linear waste stream and are disposed of through high-temperature
incineration, with no recovery of materials or components.

\subsubsection{Circular Supply Chain Strategy 1: Remanufactured Single-use Laparoscopic Scissors}
Under this CE-based strategy, a remanufacturing loop is integrated into the LE model (Figure~\ref{fig:conceptual}b). In this configuration, instead of  incineration, used SU instruments are collected and transported to a reprocessing facility where they undergo inspection and remanufacturing, and are subsequently dispatched back to hospitals for additional use cycles.

\subsubsection{Circular Supply Chain Strategy 2: Hybrid Laparoscopic Scissors}

Hybrid laparoscopic scissors are modular instruments comprising a reusable handle designed for multiple use cycles and a disposable SU blade. Due to their design, the hybrid supply chain involves both a forward flow and a circular reverse loop (Figure~\ref{fig:conceptual}c).

Initially, both components are manufactured, stored, shipped to the DC, and dispatched to hospitals. At the hospital, the instrument is assembled prior to the surgical procedure. After use, the instrument is disassembled. The disposable blade enters a linear waste stream for disposal, while the reusable handle enters a circular loop requiring cleaning, decontamination, and inspection. Handles requiring repair are sent to an external repair centre, re-inspected, and returned to inventory. Handles not requiring repair are placed directly back into inventory, to be assembled with a new blade for the next use cycle.

\subsubsection{Circular Supply Chain Strategy 3: Fully Reusable Laparoscopic Scissors}

Fully RU laparoscopic scissors are designed entirely for multiple use cycles. As the entire device is circular, the supply chain flow operates similarly to the reverse loop of the hybrid model (Figure~\ref{fig:conceptual}d), necessitating disassembly, cleaning, decontamination, and inspection after each procedure.
The key distinction is the absence of a linear waste stream. Following disassembly, both the handle and the blade are checked and, if required, sent to the external repair centre before being returned  for subsequent use cycles.

In this context, the overarching aim of this study is to evaluate the environmental, financial, and operational impacts of transitioning to these circular laparoscopic scissors with an NHS supply chain and to identify the conditions under which each circular strategy becomes viable relative to the existing LE system. To address this aim, the simulation model is developed to answer the following research questions (RQs):

\begin{enumerate}

    \item What is the effective number of cycles required to achieve a reduction in emissions and costs for CE remanufactured SU laparoscopic scissors?
    \item What is the effective number of cycles required to achieve a reduction in emissions and costs for CE hybrid laparoscopic scissors?
    \item What is the effective number of cycles required to achieve a reduction in emissions and costs for CE fully reusable laparoscopic scissors?
    \item What is the optimal hospital re-order point to ensure a 100\% service level (i.e., ensuring adequate stock is always available in line with demand for surgical procedures)?
    \item What is the target cost of the hybrid laparoscopic scissor blade required to achieve financial parity with the base LE scenario?
\end{enumerate}

\subsection{Modelling Approach and Rationale}
\label{sec:methodology-approach}

We adopt a hybrid ABS-DES simulation as we consider that it can represent in combination the supply and recovery chain (DES model) and autonomous behaviour of individual supply chain members (ABS model) \citep{Brailsford2019}. More specifically, the ABS component defines the structure of the supply chain. Each SC member is modelled as an agent with its own attributes, inventory policy, and decision rules. Further, agents are connected through a GIS-based network that determines transport distances and lead times between them. The DES component defines the behaviour within each agent, representing the internal process flows through which instruments are produced, stored, transported, used, decontaminated, repaired, and reassembled. The two components interact in both directions. Within each agent, a statechart monitors the inventory level maintained by that agent's DES process flow and, when the level falls below the reorder point, triggers the generation of an order. The order is transmitted to the upstream agent selected by the receiving agent's decision rules, where it enters that agent's process flow and initiates dispatch or production. Instruments dispatched in response travel across the agent network, with transit times determined by the distance between agents, and re-enter the receiving agent's process flow on arrival. Agent-level decisions therefore determine when and to whom material flows are initiated, while the DES process flows determine how long each operation takes, what resources it consumes, and what cost and emissions it generates.

No single simulation paradigm satisfies all of these requirements. System dynamics  can represent the supply and recovery system effectively at an aggregate level \cite{Randersed1997, Tako2012} but, it cannot track individual instruments through successive use cycles, nor can it represent stochastic events at process-level, capturing the heterogeneity of individual SC members. For this reason, SD is not considered suitable for our model. DES can represent detailed operational processes for each supply chain member and is well-suited for tracking entities through event-driven systems \citep{Goldsman2015}, but it pre-specifies the interactions between SC members and is less suitable in representing the autonomous, rule-based decision-making of multiple SC members. ABS can represent autonomous agents and emergent inter-organisational interactions \citep{Macal2009} but lacks the natural constructs for representing detailed sequential process flows within each member.

\begin{uselscape}
\begin{figure}
    \centering
    \includegraphics[width=\linewidth]{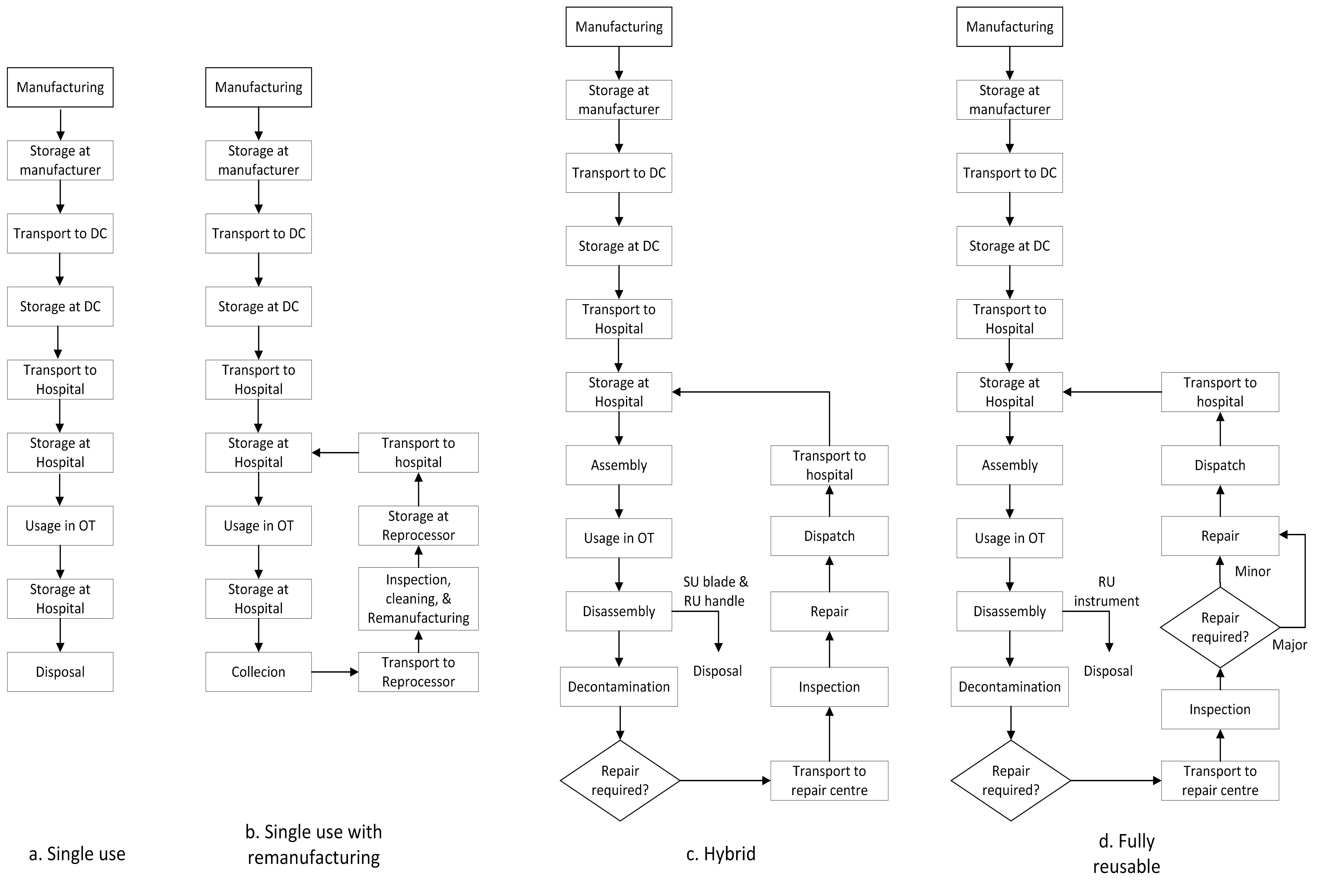}
    \caption{Process flows for (a) linear single-use, (b) remanufactured single-use, (c) hybrid, and (d) fully reusable laparoscopic scissors.}
    \label{fig:conceptual}
\end{figure}
\end{uselscape}

\subsection{The Conceptual Model}
\label{sec:methodology-overview}

\begin{figure}[h]
    \centering
    \includegraphics[width=\linewidth]{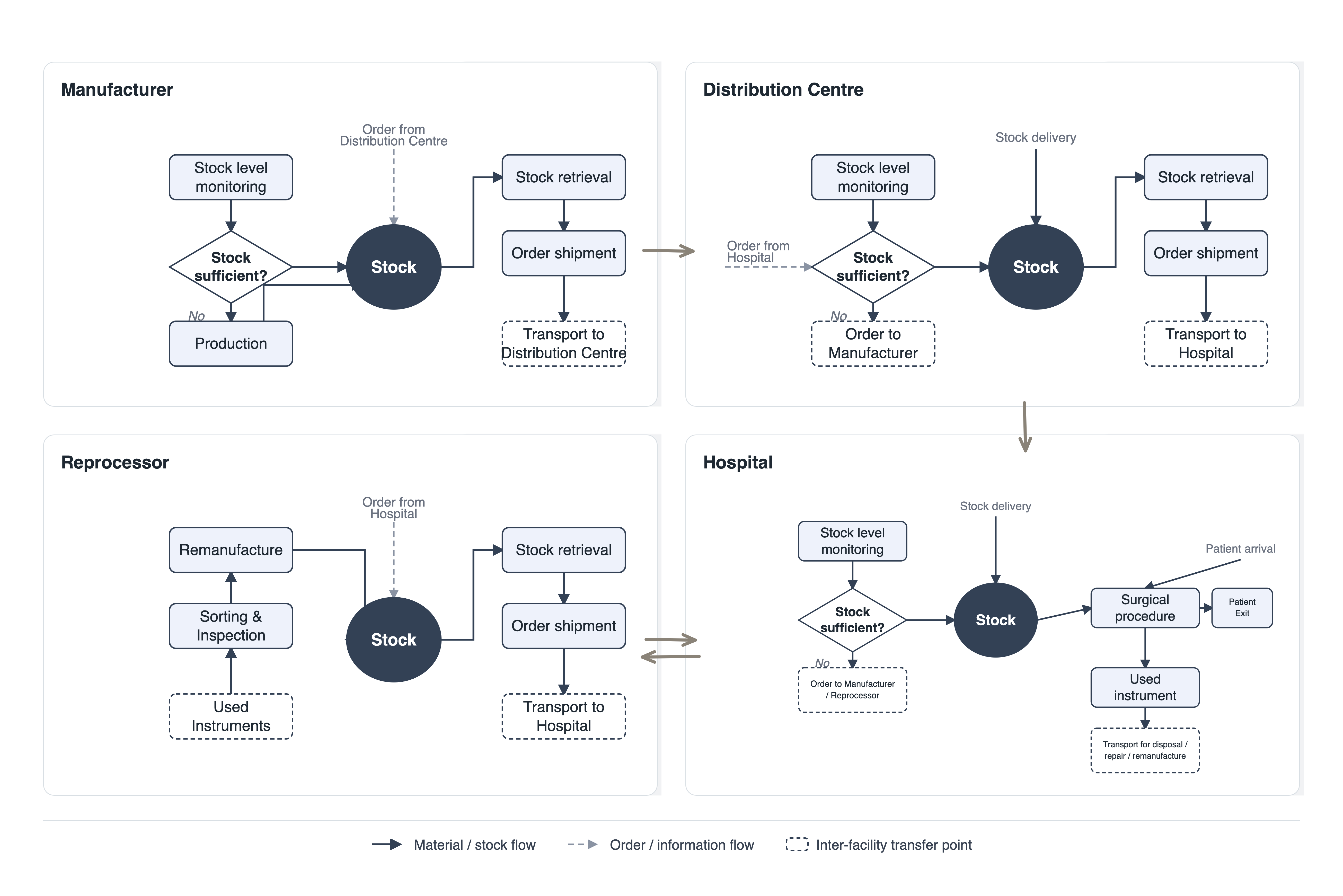}
    \caption{High-level representation of the simulation model logic depicting the flow of instruments across the supply and recovery chain.}
    \label{fig:conceptual-diagram}
\end{figure}

The simulation model represents a four-echelon supply and recovery chain for laparoscopic scissors, comprising: manufacturer, distribution centre, hospital, and reprocessor (see Figure \ref{fig:conceptual-diagram}). The supply chain (representing the forward flow of scissors) captures product movement from production to use, while the recovery chain (reverse flow) represents circular processes, including reuse, repair, and remanufacturing. The network modelled in this study comprises one manufacturer, one distribution centre, nine hospitals, and one reprocessing centre, reflecting the structure of a regional NHS supply network.

Supply chain members are modelled as autonomous agents embedded within a network structure. In addition to these primary agents, auxiliary agents such as transport trucks and orders are included to facilitate the transport of products and flow of information within the system, respectively. These auxiliary agents are passive and do not exhibit independent decision-making behaviour.

Hospitals act as the focal agents in the model, which generate demand that is driven by patients' arrival for surgery into hospital. As laparascopic scissors are used, inventory levels decline, triggering replenishment orders when stock falls below a predefined reorder threshold. Orders are sent to the distribution centre, which fulfils demand and replenishes its inventory by placing orders with the manufacturer. After being used, laparoscopic scissors are routed depending on the specific circular strategy adopted, which could be either disposal (if singe use), cleaning or repair (if hybrid or fully RU), or reprocessing (if SU with remanufacturing) as shown in Figure \ref{fig:conceptual}. 

The manufacturer operates under a make-to-stock policy, fulfilling orders from inventory and initiating production when stock levels reach a minimum threshold. The operational logic implemented for each supply chain member is illustrated in Figure \ref{fig:conceptual-diagram}.

The model is implemented in AnyLogic and initialised and parameterised from a frontend Microsoft Excel spreadsheet. Key inputs — including geographic coordinates for GIS-based placement, network configuration, demand profiles, and cost and emissions parameters — are imported at the initialisation stage. This approach enables flexible reconfiguration of the supply chain structure without modification of the underlying model logic, which can be adopted for any other type of small medical devices flowing through a similar network configuration other than laparascopic scissors considered here.

The operations and model logic for each member (manufacturer, distribution centre, hospital, and reprocessor) in the supply and recovery chain are described in detail in Appendix \ref{app:methodology_member}.

\subsection{Performance Indicators}
\label{sec:kpis}

Two primary types of performance indicator are used in this study, environmental and financial. Additionally, operational performance indicators  are used to assess how operational demands are fulfilled due to the adoption of circular products.
 
\textit{Environmental performance} is measured through carbon dioxide
equivalent (CO\textsubscript{2}-eq) emissions and virgin raw material consumption. Emissions encompass generation across the laparoscopic scissors lifecycle, including manufacturing and material extraction, transportation by truck and ship, disposal, and where applicable, decontamination, sterilisation, and repair cycles.
Material consumption is measured through the mass of virgin metal and plastic required to manufacture new instruments over the simulation period. Each instrument type is assigned a total weight and a material composition and the consumption is computed by multiplying the number of newly manufactured instruments, or components (for hybrid), by their respective metal and plastic mass.
 
\textit{Financial performance} is measured through total lifecycle
supply chain costs.
Purchase or procurement costs are assumed to already include manufacturing, distribution, and storage costs. Disposal costs are calculated separately. For CE scenarios, costs additionally include staffing for laparoscopic scissors handling, decontamination, repair, and courier services for transport
to the repair centre.

\textit{Operational performance.} For hybrid and fully reusable systems, operational performance is evaluated using decontamination throughput (volume of instruments processed per month), cumulative staff handling time (man-hours required for handling including retrieval, storage, and assembly).

\subsection{Main Model Assumptions}
\label{sec:assumptions}

To model the supply and recovery network for laparoscopic scissors the following key assumptions are made. Additional assumptions are provided in Appendix \ref{app:sensitivity}. 
 
\begin{itemize}
 
  \item  The SC comprises one manufacturer, one distribution centre, nine  hospitals, and a single reprocessor, configured to represent a small region in the UK.
  
  \item One laparoscopic scissor is required per procedure, and all procedures require laparoscopic scissors.
  
  \item Demand for laparoscopic scissors at each hospital is assumed to remain stationary over the simulation period, based on patient arrival rates.
  
  \item All cost parameters are assumed to remain constant over the simulation period. Inflation, changes in NHS pay scales, and future instrument pricing are not modelled.

   \item The repair process is not modelled as a sequence of operations; rather, it is abstracted as a time delay that holds the laparoscopic scissors for a specified duration and incurs an associated repair cost.

   \item SU laparoscopic scissors are remanufactured (re-instated to a new-like condition) at reprocessor agent and subsequently CE-marked to indicate fitness for use.

   \item A 10\% probability of instrument failure per use cycle is
  assumed; in the event of failure, a replacement instrument is
  used. 
 
\end{itemize}

\subsection{Simulation Input Data}
\label{sec:inputdata}

To parameterise the model we use data collected from different sources, including published literature, UK government reports, and industry-informed estimates provided by subject matter experts.
A central challenge in modelling a CE transition is the limited availability of data for commercially sensitive data, such as laparoscopic scissors costs and carbon footprints across market variants.
To address this and to account for the inherent variability of real-world systems, key parameters are represented using statistical distributions rather than deterministic point estimates.
 
For parameters derived from the literature as single point estimates, stochasticity was introduced by fitting a Triangular distribution a $\pm20\%$ range around the mean. For example, manufacturing emissions for the reusable handle of a hybrid instrument used the mean value of 635~gCO$_2$-eq \cite{rizan2022environmental} as the mode, with minimum and maximum set at 508 and 762, respectively.
This is a well-established practice in simulation modelling for representing uncertainty when empirical data for a full distribution is unavailable \citep{Biller2010}.
For parameters where industry experts provided a range of values, Uniform distributions were used to reflect equal likelihood across that range.

A detailed description of the financial, environmental, and operational
parameters, together with full distributions, values, and sources, is
provided in Appendix~\ref{app:inputdata}.

\subsection{Scenarios Modelled}
\label{sec:scenarios}

For the scenario analysis, three CE
strategies are evaluated against a LE base scenario, as detailed in
Table~\ref{tab:scenarios}.
The base scenario represents current practice: SU scissors with
high-temperature incineration.
 
Scenario~1 includes two sub-scenarios related to remanufacturing of SU
scissors.
Scenario~1a (controlled adoption) evaluates remanufacturing with a
fixed 50\% adoption rate at hospitals, with use cycles varied between
1 and~5.
Scenario~1b (prioritised adoption) builds on Scenario~1a, but
remanufactured laparoscopic scissors are prioritised over new ones when
both are available, with use cycles again varied between 1 and~5.
 
Scenario~2 assesses the adoption of hybrid scissors with handle use
cycles varied between 1 and~70.
Reusable handles are paired with disposable SU blades, and a non-zero
probability of instrument malfunction is included, necessitating
repair.
 
Scenario~3 assesses fully RU scissors with use cycles varied between 1
and~70, with a repair option included.

\begin{table}[htbp]
\centering
\caption{Simulation scenarios and factors varied.}
\label{tab:scenarios}
\small
\begin{threeparttable}

  \begin{tabular}{p{2cm} p{8cm} p{3cm}}

  \toprule
  {Instrument type}
    & {Scenario}
    & {Factors varied} \\
  \midrule

  \multirow{9}{2.8cm}{SU scissors}
    & {Base case:} SU instruments with
      incineration as the EoL
      option.
    & --- \\[18pt]

    & {Scenario 1a} (Controlled adoption):
     Remanufacturing at fixed 50\% hospital adoption rate; 
     new and remanufactured used in equal proportion..
    & Use cycles varied between 1 and 5. \\[18pt]

    & {Scenario 1b} (Prioritised adoption):
      Remanufacturing of SU scissors with
      prioritised adoption of remanufactured
      scissors.
    & Use cycles varied between 1 and 5. \\[6pt]

  \midrule

  Hybrid scissors
    & {Scenario 2:}
      Adoption of hybrid scissors comprising a
      reusable handle paired with a disposable
      SU blade. Handle reuse and repair included.
    & Use cycles of the reusable handle varied
      between 1 and 70. \\[6pt]

  \midrule

  Fully RU scissors
    & {Scenario 3:}
      Adoption of fully reusable scissors with
      all components entering a circular loop
      after each use cycle. Repair included.
    & Use cycles varied between 1 and 70. \\[2pt]

  \bottomrule
  \end{tabular}

\end{threeparttable}
\end{table}

\subsection{Model Validation}
\label{sec:validation}

As this study models prospective CE systems that do not yet exist, formal validation against historical real-world data is not possible. In such cases, the use of multiple complementary validation techniques is recommended \cite{sargent2010verification}. Accordingly, a multi-faceted validation approach was adopted.

First, conceptual validation was conducted through close collaboration with industry experts with domain knowledge of medical device supply chains. The model structure, assumptions, and system boundaries were iteratively developed and reviewed to ensure they adequately represent the real-world system \cite{sargent2010verification}.

Secondly, extensive verification and structural validation testing were carried out. This included executing the model under simplified deterministic conditions to ensure that outputs were as expected and consistent with expected system behaviour, thereby confirming correct implementation of the model logic \cite{balci1997verification}.

Third, face validation was performed by presenting model outputs and behaviour to domain experts in reprocessing and hospital supply chain, who confirmed that the results were plausible and consistent with their real world experience in practice \cite{sargent2010verification, robinson2025simulation}.

\section{Model Results and Findings}
\label{sec:results}

Simulation experiments were conducted using the AnyLogic software. The model was run over a 10-year period following a warm-up period to ensure steady-state conditions.

The number of replications was determined using the graphical method proposed by \cite{Robinson1994}. The approach involves visually inspecting the plot for key model outputs and identifying where a steady state is achieved. In our case, we plotted the cumulative mean of the number of laparoscopic scissors used against the number of replications. A steady state was observed after approximately 20 replications, with the half-width of the 99\% confidence interval falling below 1\% of the mean. A total of 30 replications were therefore used to ensure statistical robustness.

Results are presented first for the LE base scenario, followed by CE scenarios 1 (a and b), 2 and 3, which are compared against this base scenario.

\subsection{LE Base Scenario: Single-Use Laparoscopic Scissors}
\label{sec:results-LE}

Table \ref{tab:basecase} summarises the simulation outcomes for the base case scenario, reporting mean values and 95\% confidence intervals.

Over the 10-year simulation horizon, approximately 140,447 new SU scissors were used and 15,583 failed. Across the nine hospitals, an average stock level of 144.7 scissors was maintained. The total supply chain cost was approximately £3.96 million, while total emissions reached 130 tonnes of CO$_2$-eq.

These results set the base (LE) scenario performance against which all other CE scenarios are evaluated.

\begin{table}[htbp]
\centering
\caption{Simulation model outcomes for the base case scenario (LE,
         single-use laparoscopic scissors with incineration).}
\label{tab:basecase}

\begin{threeparttable}
\small
  \begin{tabular}{p{6cm} p{3cm} p{3.5cm}}
  \toprule
  Item & Mean & {95\% CI [LL, UL]} \\
  \midrule
  New laparoscopic scissors utilised (units) & 140,447 & [140,291;\ 140,602] \\[4pt]

  Laparoscopic scissors failed (units)     & 15,583     & [15,519;\ 15,648] \\[4pt]

  Average stock in hospitals (units)     & 144.7     & [140.25;\ 149.15] \\[4pt]

  Total supply chain cost (£\,GBP)     & 3,956,959     & [3,951,471;\ 3,962,447] \\[4pt]

  Supply chain emissions (kg CO\textsubscript{2}-eq)     & 130,211     & [130,012;\ 130,409] \\[2pt]

  \bottomrule
  \end{tabular}
  \end{threeparttable}
  \end{table}

\subsection{CE Scenario 1: Single-Use Laparoscopic Scissors with Remanufacturing}
\label{sec:results-CE1}
The introduction of remanufacturing as a CE strategy yields significant benefits relative to the base case. Results for Scenarios 1a (controlled adoption, 50\% rate) and 1b (prioritised adoption) are summarised in tables \ref{tab:results_1a} and \ref{tab:results_1b} in Appendix \ref{app:add-results}, respectively.

\subsubsection{Environmental Outcomes}
\label{sec:results-CE1-env}

The introduction of remanufacturing achieves substantial reductions in emissions compared to the base scenario. However, the magnitude of these reductions depends on the strategy adopted.

The controlled adoption (Scenario 1a), results in Table \ref{tab:results_1a}, show that extending the life to just two use cycles reduces total CO\textsubscript{2} emissions by approximately 33\%, from 130.2 to 88.0 tonnes of CO\textsubscript{2}-eq. However, the strategy yields diminishing returns beyond this point, as extending laparoscopic scissor life from two to five cycles achieves only a further 9.3\% reduction and about 39\% against the base scenario. 
These diminishing marginal gains are attributable to the fixed adoption rate which keep procurement volumes of new instruments largely constant regardless of use cycles, leaving upstream emissions unchanged. Despite the small magnitude of these gains, a Tukey pairwise comparison \citep{Keselman1977} confirms that the incremental reductions between consecutive cycles remain statistically significant at the 5\% level (Table \ref{tab:tukey_1a_emissions}, Appendix \ref{app:LE-results}).

In contrast, the prioritised adoption strategy (Scenario 1b) produces greater environmental benefits. The specific reduction in emissions is considerably higher, as shown in Table \ref{tab:results_1b}. Emissions decrease by approximately 37\% at two use cycles, and by over 55\% at five cycles.
This sustained decrease stems from replacing new with remanufactured scissors. A Tukey pairwise comparison confirms that these incremental reductions remain statistically significant at the 5\% level across all use cycles (Table~\ref{tab:tukey_1b_emissions}, Appendix~\ref{app:LE-results}).

\subsubsection{Financial Outcomes}
\label{sec:results-CE1-finan}

Both remanufacturing strategies result in significant cost reductions compared to the base scenario.

For Scenario 1a, the controlled 50\% adoption reduces costs by approximately 25\% at two use cycles. 
However, further increases in use cycles yield only marginal gains.
This occurs because the majority of savings are realised at the first reuse, with further life extensions yielding only reductions in disposal costs. A Tukey pairwise comparison (Table~\ref{tab:tukey_1a_cost} in Appendix~\ref{app:add-results}) confirms that the only statistically significant cost difference is between use cycles 2 and 5, implying that meaningful financial gains under controlled adoption require extending instrument life by at least three additional cycles beyond the first reuse.

In contrast, the prioritised adoption (Scenario 1b) leads to greater and sustained cost reductions. At two use cycles, costs fall by approximately 28\%, from £3.96 million to £2.85 million. Unlike Scenario 1a, further extension from 2 to 5 use cycles yields an additional cost reduction of nearly 19\%, a decrease from £2.85 million to £2.31 million, which is a total reduction of approximately 41\% against the base scenario. This is driven by the reduced procurement of new laparoscopic scissors. A Tukey pairwise comparison (Table~\ref{tab:tukey_1b_cost} in Appendix~\ref{app:LE-results}) confirms that the difference in cost between every consecutive use cycle is statistically significant (p-value = 0), implying that under the prioritised strategy, every extension of instrument life provides a significant financial benefit.

\begin{uselscape}

\begin{table}[p]
\centering
\caption{CO\textsubscript{2} emissions (kg CO\textsubscript{2}-eq) and
         supply chain costs (£\,GBP) for the base scenario (LE)
         versus CE Scenario~1a (controlled adoption, 50\% hospital
         adoption rate) across use cycles.}
\label{tab:results_1a}
\begin{threeparttable}
\small
  \begin{tabular}{
      L{3.8cm}          
      C{3.6cm}         
      C{3.6cm}          
      C{3.6cm}          
      C{3.6cm}         
      C{3.6cm}         
    }

  \toprule
  \multirow{2}{*}{{SC Member / Item}}
    & {Base case}
    & \multicolumn{4}{c}{{Scenario 1a --- Controlled adoption (50\%)}} \\
  \cmidrule(lr){3-6}
    & {1 cycle}
    & {2 cycles}
    & {3 cycles}
    & {4 cycles}
    & {5 cycles} \\
  \midrule

  \multicolumn{6}{l}{\textit{Emissions in kg CO\textsubscript{2}-eq}} \\[3pt]

  Manufacturer
    & \makecell{106,070 \\ {[}104,908;\ 107,063{]}}
    & \makecell{51,206 \\ {[}51,116;\ 51,296{]}}
    & \makecell{51,104 \\ {[}50,036;\ 52,172{]}}
    & \makecell{51,187 \\ {[}51,099;\ 51,274{]}}
    & \makecell{51,195 \\ {[}51,092;\ 51,298{]}} \\[8pt]

  Distribution centre (transport)
    & \makecell{175 \\ {[}173.7;\ 176.6{]}}
    & \makecell{87.97 \\ {[}87.17;\ 88.76{]}}
    & \makecell{87.90 \\ {[}86.81;\ 88.99{]}}
    & \makecell{90.71 \\ {[}83.76;\ 97.65{]}}
    & \makecell{87.79 \\ {[}86.62;\ 88.96{]}} \\[8pt]

  Hospital (disposal)
    & \makecell{24,030 \\ {[}23,875;\ 24,224{]}}
    & ---
    & ---
    & ---
    & --- \\[8pt]

  Reprocessor
    & ---
    & \makecell{36,731 \\ {[}36,426;\ 37,036{]}}
    & \makecell{31,296 \\ {[}30,996;\ 31,596{]}}
    & \makecell{29,471 \\ {[}29,244;\ 29,698{]}}
    & \makecell{28,550 \\ {[}28,364;\ 28,736{]}} \\[8pt]

  {Total}
    & \makecell{{130,211} \\ {{[}129,124;\ 131,297{]}}}
    & \makecell{{88,025} \\ {{[}87,693;\ 88,358{]}}}
    & \makecell{{82,488} \\ {{[}81,444;\ 83,532{]}}}
    & \makecell{{80,749} \\ {{[}80,489;\ 81,009{]}}}
    & \makecell{{79,833} \\ {{[}79,667;\ 79,999{]}}} \\[4pt]

  \midrule
  \multicolumn{6}{l}{\textit{Costs in £\,GBP}} \\[3pt]

  New laparoscopic scissors
    & \makecell{3,950,767 \\ {[}3,920,753;\ 3,980,782{]}}
    & \makecell{1,977,114 \\ {[}1,962,723;\ 1,991,505{]}}
    & \makecell{1,973,765 \\ {[}1,956,632;\ 1,990,897{]}}
    & \makecell{1,975,142 \\ {[}1,959,173;\ 1,991,112{]}}
    & \makecell{1,973,033 \\ {[}1,955,123;\ 1,990,943{]}} \\[8pt]

  Remanufactured laparoscopic scissors
    & ---
    & \makecell{981,247 \\ {[}972,091;\ 990,404{]}}
    & \makecell{980,433 \\ {[}970,602;\ 990,264{]}\tnote{a}}
    & \makecell{979,648 \\ {[}972,206;\ 987,090{]}}
    & \makecell{978,998 \\ {[}972,512;\ 985,485{]}} \\[8pt]

  Disposal
    & \makecell{6,192 \\ {[}6,147;\ 6,237{]}}
    & \makecell{2,788 \\ {[}2,767;\ 2,810{]}}
    & \makecell{1,393 \\ {[}1,380;\ 1,407{]}}
    & \makecell{928 \\ {[}922;\ 934{]}}
    & \makecell{697 \\ {[}692;\ 701{]}} \\[8pt]

  {Total}
    & \makecell{{3,956,959} \\ {{[}3,926,900;\ 3,987,019{]}}}
    & \makecell{{2,961,150} \\ {{[}2,945,103;\ 2,977,196{]}}}
    & \makecell{{2,955,591} \\ {{[}2,938,363;\ 2,972,818{]}}}
    & \makecell{{2,955,718} \\ {{[}2,937,269;\ 2,974,168{]}}}
    & \makecell{{2,952,728} \\ {{[}2,932,925;\ 2,972,531{]}}} \\[4pt]

  \bottomrule
  \end{tabular}

\end{threeparttable}
\end{table}

\begin{table}[p]
\centering
\caption{CO\textsubscript{2} emissions (kg CO\textsubscript{2}-eq) and
         supply chain costs (£\,GBP) for the base Linear Economy scenario
         versus CE Scenario~1b (prioritised adoption) across use cycles 2--5.
         Values reported as mean [95\% CI: LL;\ UL] over 30 replications.}
\label{tab:results_1b}
\begin{threeparttable}
\small
  \begin{tabular}{
      L{3.8cm}
      C{3.6cm}
      C{3.6cm}
      C{3.6cm}
      C{3.6cm}
      C{3.6cm}
    }

  \toprule
  \multirow{2}{*}{{SC Member / Item}}
    & {Base case}
    & \multicolumn{4}{c}{{Scenario 1b --- Prioritised adoption}} \\
  \cmidrule(lr){3-6}
    & {1 cycle}
    & {2 cycles}
    & {3 cycles}
    & {4 cycles}
    & {5 cycles} \\
  \midrule

  \multicolumn{6}{l}{\textit{Emissions in kg CO\textsubscript{2}-eq}} \\[3pt]

  Manufacturer
    & \makecell{106,070 \\ {[}104,908;\ 107,063{]}}
    & \makecell{42,670 \\ {[}42,580;\ 42,760{]}}
    & \makecell{25,589 \\ {[}25,549;\ 25,630{]}}
    & \makecell{17,057 \\ {[}17,010;\ 17,104{]}}
    & \makecell{11,375 \\ {[}11,328;\ 11,422{]}} \\[8pt]

  Distribution centre (transport)
    & \makecell{175 \\ {[}173.7;\ 176.6{]}}
    & \makecell{76.12 \\ {[}75.42;\ 76.81{]}}
    & \makecell{50.29 \\ {[}49.61;\ 50.98{]}}
    & \makecell{37.38 \\ {[}36.63;\ 38.14{]}}
    & \makecell{29.51 \\ {[}28.65;\ 30.37{]}} \\[8pt]

  Hospital (disposal)
    & \makecell{24,030 \\ {[}23,875;\ 24,224{]}}
    & ---
    & ---
    & ---
    & --- \\[8pt]

  Reprocessor
    & ---
    & \makecell{39,314 \\ {[}39,095;\ 39,535{]}}
    & \makecell{43,470 \\ {[}43,383;\ 43,658{]}}
    & \makecell{45,521 \\ {[}45,268;\ 45,773{]}}
    & \makecell{46,747 \\ {[}46,494;\ 46,999{]}} \\[8pt]

  {Total}
    & \makecell{{130,211} \\ {{[}129,124;\ 131,297{]}}}
    & \makecell{{82,061} \\ {{[}81,823;\ 82,298{]}}}
    & \makecell{{69,110} \\ {{[}68,922;\ 69,298{]}}}
    & \makecell{{62,615} \\ {{[}62,358;\ 62,873{]}}}
    & \makecell{{58,152} \\ {{[}57,900;\ 58,403{]}}} \\[4pt]

  \midrule
  \multicolumn{6}{l}{\textit{Costs in £\,GBP}} \\[3pt]

  New laparoscopic scissors
    & \makecell{3,950,767 \\ {[}3,920,753;\ 3,980,782{]}}
    & \makecell{1,766,570 \\ {[}1,758,044;\ 1,775,096{]}}
    & \makecell{1,174,508 \\ {[}1,168,256;\ 1,180,760{]}}
    & \makecell{877,405 \\ {[}872,427;\ 882,383{]}}
    & \makecell{697,814 \\ {[}693,382;\ 702,246{]}} \\[8pt]

  Remanufactured laparoscopic scissors
    & ---
    & \makecell{1,080,386 \\ {[}1,073,628;\ 1,087,143{]}}
    & \makecell{1,377,898 \\ {[}1,371,899;\ 1,383,898{]}}
    & \makecell{1,525,513 \\ {[}1,516,690;\ 1,534,336{]}}
    & \makecell{1,614,201 \\ {[}1,604,877;\ 1,623,525{]}} \\[8pt]

  Disposal
    & \makecell{6,192 \\ {[}6,147;\ 6,237{]}}
    & \makecell{2,786 \\ {[}2,773;\ 2,799{]}}
    & \makecell{1,859 \\ {[}1,848;\ 1,869{]}}
    & \makecell{1,393 \\ {[}1,386;\ 1,401{]}}
    & \makecell{1,114 \\ {[}1,108;\ 1,119{]}} \\[8pt]

  {Total}
    & \makecell{{3,956,959} \\ {{[}3,926,900;\ 3,987,019{]}}}
    & \makecell{{2,849,743} \\ {{[}2,834,911;\ 2,864,575{]}}}
    & \makecell{{2,554,265} \\ {{[}2,542,717;\ 2,565,813{]}}}
    & \makecell{{2,404,312} \\ {{[}2,391,292;\ 2,417,332{]}}}
    & \makecell{{2,313,129} \\ {{[}2,299,808;\ 2,326,459{]}}} \\[4pt]

  \bottomrule
  \end{tabular}

\end{threeparttable}
\end{table}

\end{uselscape}

\subsection{CE Scenario 2: Hybrid Laparoscopic Scissors}
\label{sec:results-CE-2}

\subsubsection{Environmental Outcomes}
\label{sec:results-CE2-env}

\begin{figure}
    \centering
    \includegraphics[width=0.75\linewidth]{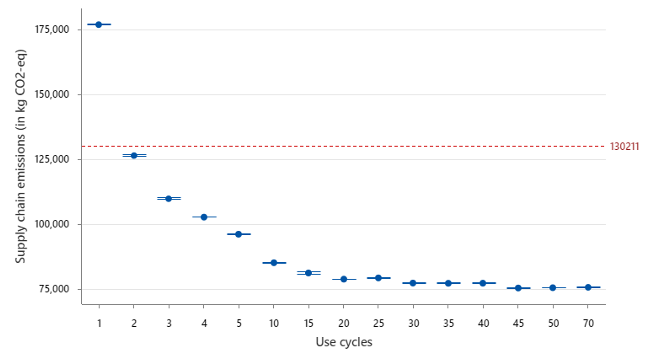}
    \caption{Supply chain emissions (mean and 95\% CI) for hybrid scissors (Scenario 2) at different use cycles against the base scenario (SU). \textit{Note: Base case indicated in red dotted line}.}
    \label{fig:hybrid-emissions}
\end{figure}

Hybrid laparoscopic scissors exhibit a non-linear environmental profile characterised by high initial carbon intensity followed by rapid improvements with reuse. 
Figure~\ref{fig:hybrid-emissions} illustrates the model results for hybrid laparoscopic scissors. The results follow a pattern of diminishing return.

At a single use cycle, carbon emissions exceed the LE base scenario by approximately 37\% driven by the higher material and manufacturing requirements associated with better quality and durable hybrid scissors. However, subsequent reuses cause a steep decline in emissions, achieving environmental breakeven at the second cycle.
Emissions continue to decrease up to 20–25 cycles, after which the rate of decline levels off and two clusters emerge. The first cluster covers use cycles 30, 35, and 40 (Group 1), while the second covers 45, 50, and 70 (Group 2). Tukey pairwise comparisons (Table~\ref{tab:tukey_sc2_emissions} in Appendix~\ref{app:ce-2-results}) confirm that reductions are statistically significant up to 25 cycles. The comparison also shows that the intra-cluster differences are statistically insignificant and the inter-cluster difference is statistically significant.

Next, we analyse the effect of the variation in use cycles on the number of laparoscopic scissors required and disposed of. 

\begin{figure}[htbp]
    \centering
    \includegraphics[width=0.75\linewidth]{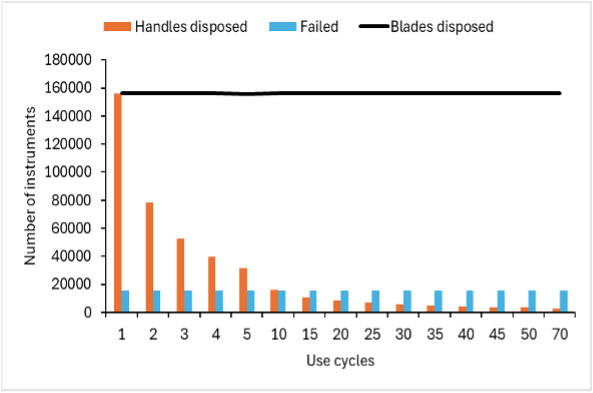}
    \caption{Number of hybrid scissors failed and number of handles and blades disposed of at different use cycles (CE Scenario 2).}
    \label{fig:hybrid-waste}
\end{figure}

Figure~\ref{fig:hybrid-waste} displays the number of scissors failed and disposed of at different use cycles. The number of SU blades disposed of remains constant at approximately 156,000 units — equivalent to the combined number of laparoscopic scissors used and failed in the base scenario, where a new blade is required for every surgical procedure. In contrast, disposal of the reusable handle components decreases from over 70,000 units at two use cycles to just over 2,500 at 70 cycles, achieving a reduction in waste of almost 96\%. The number of instrument failures remains stable at approximately 15,600 units across all use cycles, consistent with the fixed 10\% failure rate.

\begin{figure}[htbp]
    \centering
    \includegraphics[width=0.75\linewidth]{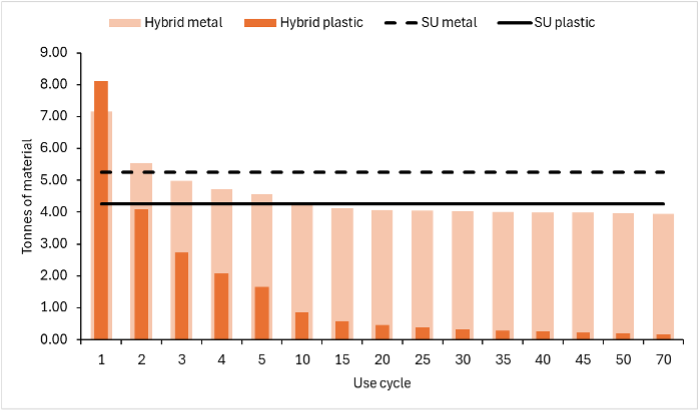}
    \caption{Total raw (virgin) material consumption (in tonnes) for hybrid scissors (CE Scenario 2) compared against base scenario (LE SU) at different use cycle.}
    \label{fig:hybrid-materials}
\end{figure}

Figure~\ref{fig:hybrid-materials} compares virgin raw material consumption for the hybrid CE strategy (hybrid laparoscopic scissors) against the base scenario (which requires more than five tonnes of metal and over four tonnes of plastic). Plastic consumption at a single use cycle is relatively high (over eight tonnes) but drops sharply, falling below the base scenario from the second use cycle onward and becoming almost negligible (less than one tonne) after twenty use cycles. Metal consumption, by contrast, reduces at a slower rate and plateaus at approximately four tonnes after fifteen use cycles — a direct consequence of the hybrid design, which requires a new metal blade at every procedure.

\subsubsection{Financial Outcomes}
\label{sec:results-CE2-finan}

\begin{figure}
    \centering
    \includegraphics[width=0.75\linewidth]{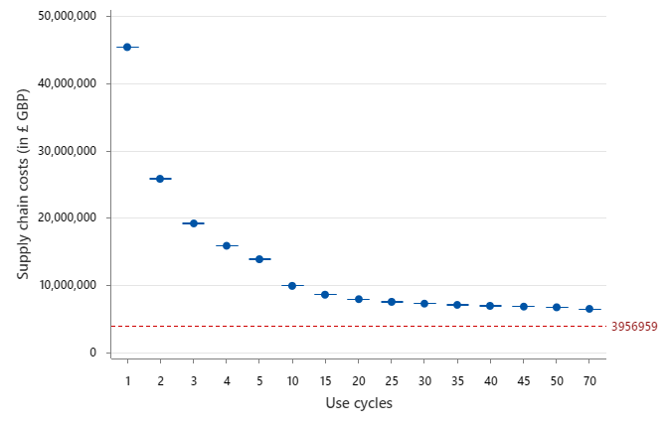}
    \caption{Supply chain costs (mean and 95\% CI) for hybrid scissors (CE Scenario 2) at different use cycles against the base scenario (SU). \textit{Note: Base scenario indicated in red dotted line.}}
    \label{fig:hybrid-cost}
\end{figure}

Figure~\ref{fig:hybrid-cost} displays total supply chain costs for hybrid laparoscopic scissors across different use cycles, benchmarked against the base scenario (approximately £3.96 million). 
Hybrid systems remain financially unfavourable across all use cycles despite reductions in cost with increased reuse.

Initial costs substantially exceed the base scenario due to the high capital investment required for reusable components. While the costs are rapidly amortised over subsequent uses, the system fails to achieve financial parity even at maximum reuse. This sustained cost premium is driven by the combined burden of the initial capital outlay and the recurring need for disposable components.
Statistical analysis (Table~\ref{tab:tukey_sc2_cost} in Appendix~\ref{app:ce-2-results}) confirms that cost differences across use cycles remain statistically significant; although marginal gains diminish at higher reuse levels.

Overall the results show that even at 70 use cycles, the hybrid system does not achieve financial breakeven with the base scenario. This is primarily due to the high recurring cost of the disposable blade. 
A more detailed examination of these cost factors is undertaken in Section~\ref{sec:sensitivity}.

\subsection{CE Scenario 3: Fully Reusable Laparoscopic Scissors}
\label{sec:results-CE-3}

\subsubsection{Environmental Outcomes}
\label{sec:results-CE3-env}

\begin{figure}
    \centering
    \includegraphics[width=0.75\linewidth]{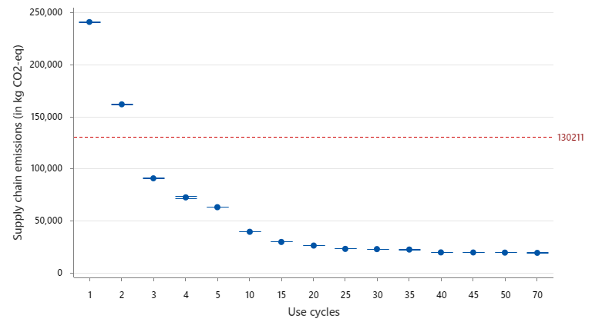}
    \caption{Total supply chain emissions in kg CO\textsubscript{2}-eq. (mean and 95\% CI) for fully reusable scissors (Scenario 3) at different use cycles against the base scenario (SU).}
    \label{fig:ru-emissions}
\end{figure}

The environmental outcomes reported by the model for fully RU laparoscopic scissors across different use cycles is depicted in Figure~\ref{fig:ru-emissions}, benchmarked against the base scenario (130.2 tonnes CO\textsubscript{2}-eq). 
These model results confirm that fully reusable scissors offer the greatest long-term environmental benefits among all tested options.

At a single use cycle, emissions are approximately 240 tonnes CO$_2$-eq, significantly exceeding the linear base scenario due to carbon-intensive material extraction and manufacturing processes associated with these higher-grade and longer lasting laparoscopic scissors.
However, emissions drop rapidly upon reuse, achieving environmental breakeven at approximately three cycles and return a 79\% reduction relative to the base scenario by 20–25 cycles.
Thereafter, the rate of reduction slows down, indicating diminishing marginal benefits as cumulative emissions from cleaning, transport, and repair begin to offset gains.

Tukey pairwise comparisons of total supply chain emissions (Table~\ref{tab:tukey_sc2_emissions}, Appendix \ref{app:ce-2-results}) confirm that use cycles 1 to 25 are mutuallly statistically different (p-value = 0), indicating substantial early-phase reductions. Beyond this point, two statistically homogeneous groups emerge. Group 1 (use cycles 25, 30, and 35) and Group 2 (use cycles 40, 45, 50, and 70), within which no significant further reduction is observed (p-value $> 0.05$). However, the two groups remain statistically different from each other, indicating a meaningful step reduction.

\begin{figure}[htbp]
    \centering
    \includegraphics[width=0.75\linewidth]{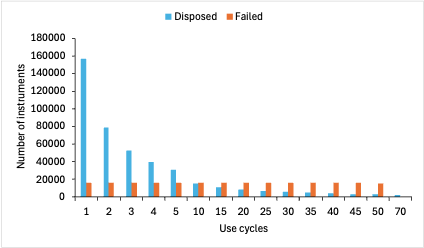}
    \caption{Number of laparoscopic scissors disposed and failed at different use cycles for fully reusable laparoscopic scissors (CE Scenario 3).}
    \label{fig:ru-waste}
\end{figure}

When adopting fully reusable laparoscopic scissors  substantial reductions in both waste generation and raw material consumption are achieved (Figure~\ref{fig:ru-waste}). In particular, total instrument disposal decreases from approximately 156,000 units in the base scenario to fewer than 2,000 units at high reuse levels, indicating that less than 2\% of the original volume is required to meet demand.

Material consumption follows a similar pattern. Although initial material requirements are higher than those of the base scenario, both metal and plastic consumption decrease rapidly with reuse. Metal usage falls below the base scenario from approximately three cycles, while plastic consumption becomes lower beyond approximately 10 cycles and approaches negligible levels at high reuse.

These results highlight the long-term efficiency gains achieved in a fully closed-loop circular economy system set up.

\begin{figure}[htbp]
    \centering
    \includegraphics[width=0.75\linewidth]{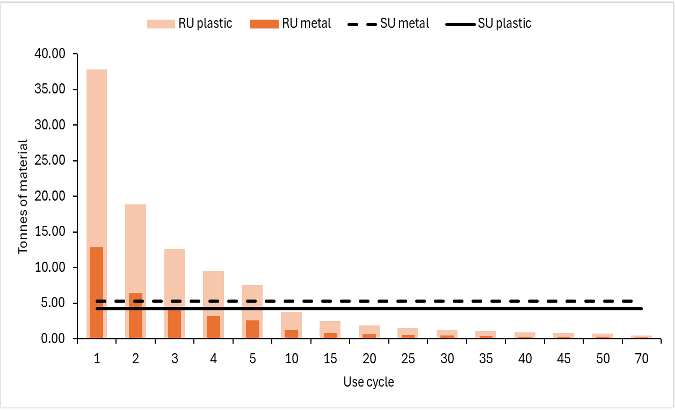}
    \caption{Amount of raw (virgin) material consumed in manufacturing fully reusable laparoscopic scissors (CE Scenario 3) against base scenario (SU) at different use cycle.}
    \label{fig:ru-material}
\end{figure}

\subsubsection{Financial Outcomes}
\label{sec:results-CE3-finan}

\begin{figure}
    \centering
    \includegraphics[width=0.75\linewidth]{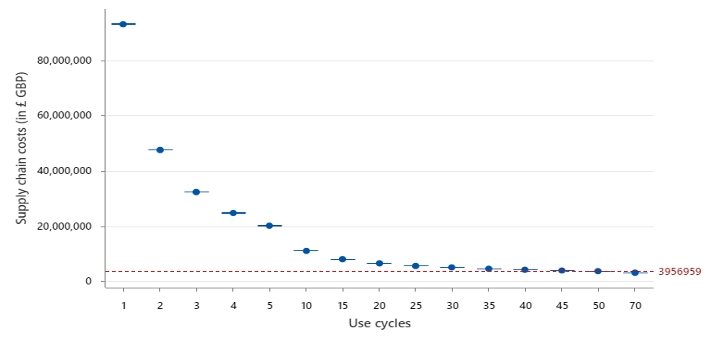}
    \caption{Supply chain costs (mean and 95\% CI) for fully reusable scissors (CE Scenario 3) at different use cycles against the base scenario (SU).}
    \label{fig:ru-cost}
\end{figure}

Figure~\ref{fig:ru-cost} presents the total supply chain costs for fully RU laparoscopic scissors across different use cycles, compared against the base scenario (approximately £3.96 million). Fully reusable systems exhibit a high initial cost but achieve financial viability at sufficiently high reuse levels.

At a single use cycle, total costs exceed £90 million due to the capital-intensive nature of reusable laparoscopic scissors. However, costs decrease rapidly with reuse as this investment is distributed over multiple cycles. A steep decline is observed up to approximately 25 use cycles, after which cost reductions become more gradual.

Financial breakeven relative to the base scenario is achieved between 45 and 50 use cycles. Thereafter, the system becomes cost-effective, with total costs decreasing to approximately £3.2 million at 70 use cycles—representing a reduction of approximately 19\% relative to the base scenario.

A simultaneous Tukey pairwise comparison (Table \ref{tab:tukey_sc3_cost}) confirms statistically significant differences (p-value $< 0.05$) across all use cycles, indicating that total costs decrease consistently with each additional reuse cycle. Unlike the hybrid CE scenario 2, the fully RU CE scenario 3 achieves financial parity with the SU base scenario. The breakeven point occurs at approximately 45-50 use cycles, which is statistically validated by independent t-tests against the base scenario (p-value $< 0.05$ at $5\%$ significance level). 

\subsubsection{Operational Performance}
\label{sec:results-CE3-oper}

The shift towards hybrid and fully RU laparoscopic scissors introduces additional processing steps at the hospital — including receiving, assembling, disassembling, cleaning, and sterilising laparoscopic scissors— after each use cycle, each of which requires dedicated staff time and resources. Since these implications are similar for both instrument types, they are presented here for the RU scenario (Scenario 3), and the findings apply equally to the hybrid case.

\begin{figure}[h]
    \centering
    \includegraphics[width=0.7\linewidth]{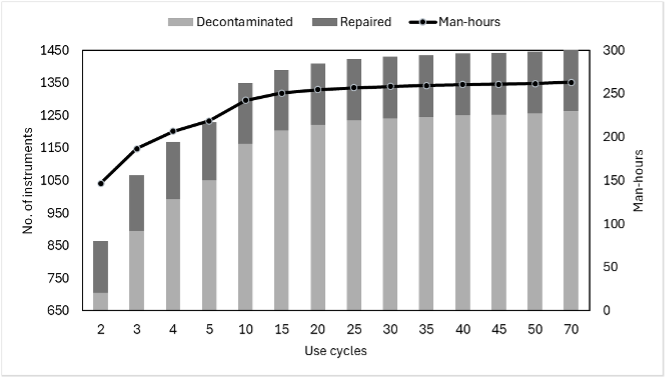}
    \caption{Decontamination and repair capacity and man-hours required at hospitals due to adoption of fully reusable laparoscopic scissors at different use cycles.}
    \label{fig:operational}
\end{figure}

Figure~\ref{fig:operational} depicts the average cumulative number of laparoscopic scissors requiring decontamination and repair, alongside the total man-hours required, as a function of use cycles. The model results demonstrate that the operational burden of a mature reusable system is considerable. At 50 cycles, the model indicates that a centralised decontamination facility serving the entire nine-hospital network must support a monthly throughput of 1,250-1,300 scissors. Furthermore, the man-hour requirement for these operational tasks levels off at approximately 260 hours at higher use cycles.

\subsection{Summary Comparison Across Scenarios}
\label{sec:cross-scenario}

Table~\ref{tab:synthesis} presents a summary of the comparison of all CE strategies, evaluated in terms of financial and environmental impact and breakeven points using as a benchmark the base LE SU scenario.  As the table illustrates, Scenario 1a (prioritised SU remanufacturing) delivers the quick returns benefits, reducing costs by about 41\% and emissions by 55\% after just two cycles. Fully RU instruments maximize environmental gains with an 86\% emissions reduction but require 45-50 use cycles to achieve financial parity. In contrast, hybrid systems achieve environmental breakeven early but fail to ever attain economic viability against the single-use baseline.
Thus, the analysis reveals that 
no single CE strategy dominates across all environmental and financial metrics.

\begin{table}[h]
\centering
\caption{Comparative performance summary of CE scenarios vs. LE base scenario.
         Emissions in tonnes CO\textsubscript{2}-eq; costs in million £\,GBP.}
\label{tab:synthesis}
\small
\begin{threeparttable}

  \begin{tabular}{
      p{3.25cm}   
      p{2.0cm}   
      p{2.5cm}   
      p{2.9cm}  
      p{2.0cm} 
    }

  \toprule
  {Scenario}
    & {Performance metric}
    & {Breakeven point}
    & {Best performance}
    & {\% change\tnote{a}} \\
  \midrule

  \multirow{2}{=}{{Base scenario (LE)}}
    & Emissions
    & N/A
    & 130.2
    & N/A \\
    & Cost
    & N/A
    & 3.96
    & N/A \\
  \midrule

  \multirow{2}{=}{{1a:} Remanufacturing (50\% adoption)}
    & Emissions
    & 2 cycles
    & 79.8 (at 5 cycles)
    & $-$38.7\% \\
    & Cost
    & 2 cycles
    & 2.95 (at 5 cycles)
    & $-$25.4\% \\
  \midrule

  \multirow{2}{=}{{1b:} Remanufacturing (prioritised adoption)}
    & Emissions
    & 2 cycles
    & 58.3 (at 5 cycles)
    & $-$55.2\% \\
    & Cost
    & 2 cycles
    & 2.32 (at 5 cycles)
    & $-$41.4\% \\
  \midrule

  \multirow{2}{=}{{2:} Hybrid laparoscopic scissors}
    & Emissions
    & 2 cycles
    & ${\sim}$75.0 (at 70 cycles)
    & $-$42.4\% \\
    & Cost
    & Never
    & ---
    & --- \\
  \midrule

  \multirow{2}{=}{{3:} Fully reusable laparoscopic scissors}
    & Emissions
    & 3 cycles
    & ${\sim}$18.0 (at 70 cycles)
    & $-$86\% \\
    & Cost
    & 45--50 cycles
    & ${\sim}$3.2 (at 70 cycles)
    & $-$19\% \\

  \bottomrule
  \end{tabular}

  \begin{tablenotes}[flushleft]\footnotesize
    \item[a] Percentage change in best performance relative to the
             base scenario. Negative values indicate improvement.
             ${\sim}$ denotes approximate values.
  \end{tablenotes}

\end{threeparttable}
\end{table}

We next enhance the analysis presented above to identify the optimal inventory levels that ensure required RU laparoscopic scissors are available to fulfill demand for surgical procedures (RQ 4) and also the cost point of blades at which hybrid laparoscopic scissors become financially viable (RQ 5).

\section{ Simulation Optimisation: Hospital Inventory Levels}
\label{sec:sim-opt}

A critical operational challenge for hospitals transitioning to CE is adjusting inventory policies to accommodate lead time variability from decontamination and repair return loops, thereby avoiding stockouts. To address this, a simulation-optimisation approach is adopted to identify the minimum reorder point ($s$ in a Min-Max policy) that a hospital with ten procedures per day must maintain to achieve a 100\% service level — defined as zero patient waiting time due to instrument unavailability. The optimisation formulation is as follows:

\begin{equation}
\min \quad Z = s \label{eq:opt_obj}
\end{equation}
Subject to:

\begin{align}
\sum_{n=1}^{N} W_n &= 0 \label{eq:opt_c1} \\
S - s &> 0 \label{eq:opt_c2} \\
S, s & > 0 \label{eq:opt_c3} 
\end{align}
where $n$ represents an arriving patient, $s$ is the reorder point, $S$ is the maximum stock level (set at six days of demand), and $W_n$ is the waiting time for the n\textsubscript{th} patient. The objective function (Equation~\ref{eq:opt_obj}) minimises the reorder point. Constraint~(\ref{eq:opt_c1}) ensures no procedure is delayed due to instrument unavailability. Constraint~(\ref{eq:opt_c2}) restricts $s$ to remain below the maximum permissible stock level
$S$. Equation~\ref{eq:opt_c3} is a non-negativity constraint.

Three scenarios with distinct instrument turnaround time profiles are defined (Table~\ref{tab:inventory_scenarios}), representing increasing levels of operational uncertainty commonly observed in practice. 

\begin{table}[htbp]
\centering
\caption{Simulation optimisation experimental scenario setup for hospital inventory
         level determination.}
\label{tab:inventory_scenarios}
\small
\begin{threeparttable}

  \begin{tabular}{p{6.5cm} p{7.1cm}}

  \toprule
  \textbf{Scenario}
    & \textbf{Turnaround time distribution} \\
  \midrule

  Scenario 3a: In-house (low variability)
    & 90\% of cases: 12 hours; \newline
      10\% of cases: 30 hours \\[6pt]

  Scenario 3b: Outsourced (deterministic)
    & Fixed 30 hours \\[6pt]

  Scenario 3c: Outsourced (high variability)
    & 50\% of cases: $\mathrm{Triangular}(20,\;30,\;40)$ hours \newline
      30\% of cases: $\mathrm{Triangular}(40,\;50,\;60)$ hours \newline
      20\% of cases: Fixed 75 hours \\[2pt]

  \bottomrule
  \end{tabular}

  \begin{tablenotes}[flushleft]\footnotesize
    \item Turnaround time is defined as the time elapsed from dispatch
          of a used instrument from the hospital to its return to
          hospital inventory following decontamination or repair.
          Triangular$(a,\;m,\;b)$ denotes a Triangular distribution
          with minimum $a$, mode $m$, and maximum $b$.
  \end{tablenotes}

\end{threeparttable}
\end{table}

\begin{figure}[h]
    \centering
    \includegraphics[width=.65\linewidth]{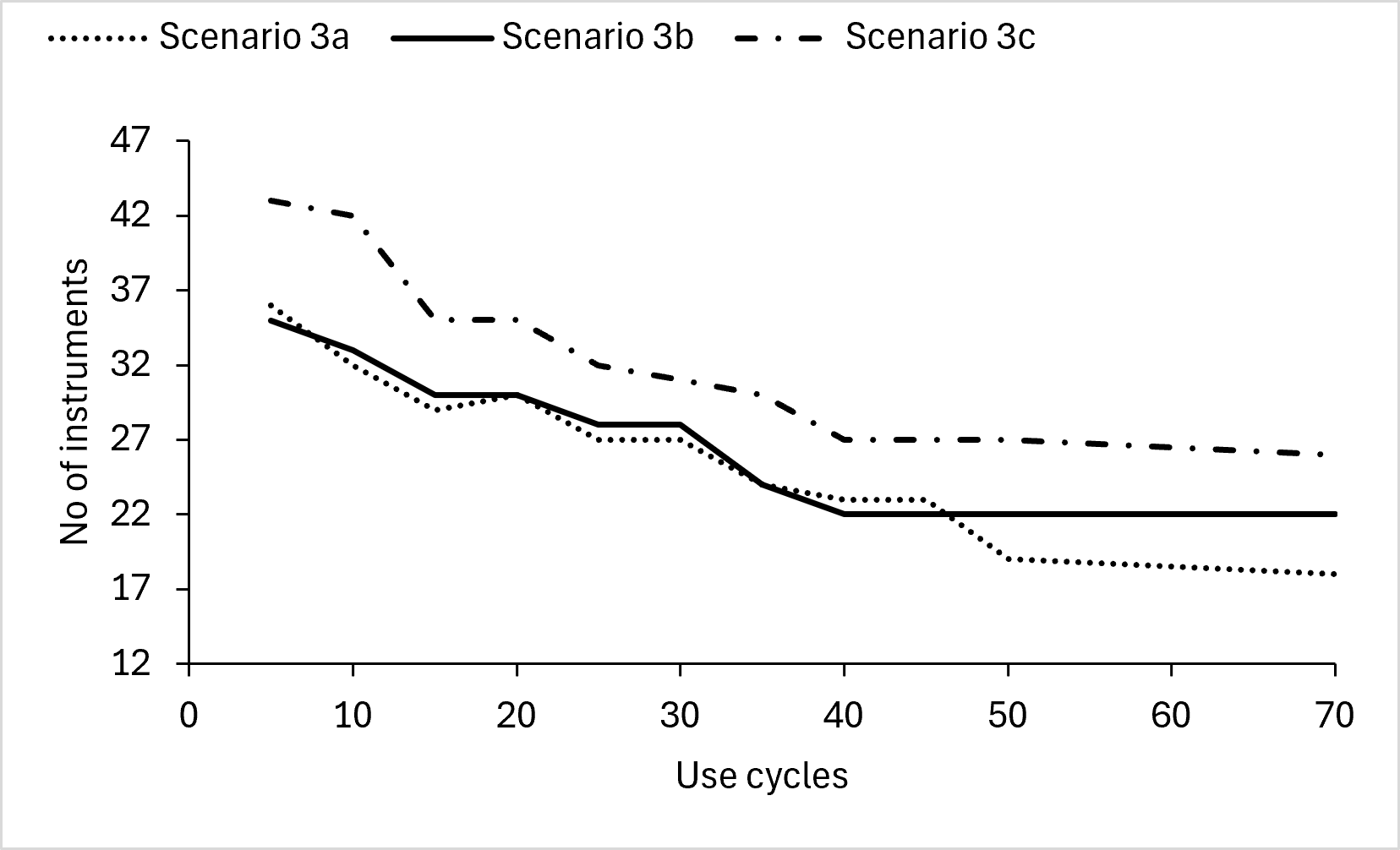}
    \caption{Optimal inventory levels by scenarios considered at different instrument use cycles.}
    \label{fig:inventory}
\end{figure}

The results are displayed in Figure~\ref{fig:inventory}. All three scenarios exhibit a similar curve profile: higher inventory is required at lower use cycles, when reliance on new stock is greater, with inventory requirements stabilising as use cycles increase and the reliance on new laparoscopic scissors diminishes. Scenario 3a (in-house) is the most operationally efficient, requiring only 22 laparoscopic scissors at 35 use cycles and 18 at 70 cycles. Scenario 3b (outsourced, deterministic) is operationally comparable to Scenario 3a, requiring only a few additional laparoscopic scissors to accommodate the slightly longer turnaround time — approximately 22 scissors at 50 use cycles. Scenario 3c (outsourced, high variability) requires considerably more safety stock due to the compounded uncertainty from both patient arrivals and instrument turnaround times. Approximately 43 laparoscopic scissors are required initially, falling to around 27 by 30 use cycles and stabilising near 25 laparoscopic scissors from 40 use cycles onwards. This translates to an additional 0.5 to 0.75 of a day's demand relative to Scenarios 3a and 3b.

\subsection{Cost Sensitivity Analysis: Hybrid Laparoscopic Scissors}
\label{sec:sensitivity}

The results in Section~\ref{sec:results-CE-2} showed that adopting hybrid  scissors (CE Scenario 2) is more expensive than the SU base LE scenario for all tested use cycles. A primary driver is the cost of the disposable blade, which is comparable to the cost of an entire SU instrument, suggesting that it may not become financially viable without a reduction in blade cost. To investigate this further, sensitivity analysis is performed to identify the blade price point at which the hybrid system achieves cost parity with the base LE (SU) scenario at three representative higher use cycles: 45, 50, and 70.

\begin{figure}
    \centering
    \includegraphics[width=0.995\linewidth]{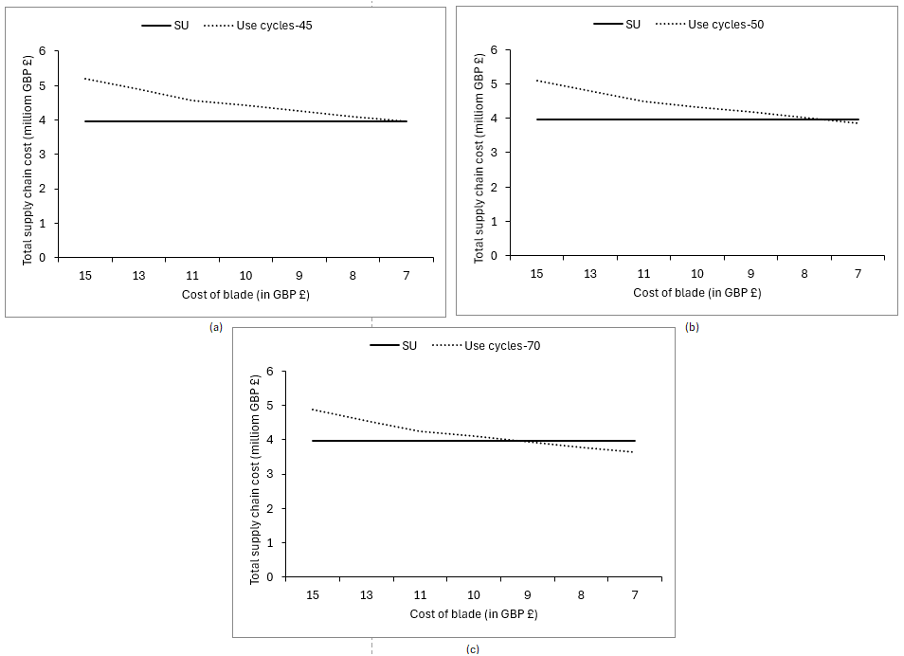}
    \caption{Cost breakeven analysis for the hybrid laparoscopic scissors at: a) 45 use cycles, b) 50 use cycles and c) 70 use cycles.}
    \label{fig:sensitivity}
\end{figure}

The results are presented in Figure~\ref{fig:sensitivity}. At 45 use cycles, cost parity is achieved when the blade cost is reduced to approximately £7; at 50 cycles, the breakeven blade cost is approximately £8; and at 70 cycles, it is approximately £9.20. However, these breakeven price points are substantially below the current minimum market price for a disposable blade — estimated at £16 based on expert opinions from industry (Table~\ref{tab:hybrid}). This indicates that achieving financial parity solely through blade cost reduction is practically unlikely under current market conditions.

To gain a better understanding of the financial drivers, a full factorial experiment evaluates the individual and combined effects of three factors — use cycles, handle cost, and blade cost — on total supply chain cost (Table~\ref{tab:factorial} in Appendix~\ref{app:sensitivity}). The main effects plot (Figure~\ref{fig:main-effect} in Appendix~\ref{app:sensitivity}) shows that all three factors have a considerable impact on supply chain costs. Use cycles is the most significant factor, followed by handle cost, with blade cost having the least — though still positive — impact. Interaction plots (Figure~\ref{fig:interaction-effect} in Appendix~\ref{app:sensitivity}) reveal no interaction between use cycles and blade cost, or between blade cost and handle cost, consistent with expectations. In contrast, an interaction exists between use cycles and handle cost, as the handle cost is a one-time procurement cost that is amortised over the number of reuse cycles, thereby diminishing in per-unit impact as use cycles increase. This suggests that the blade is procured for every procedure, it sets the floor beneath which the hybrid system's costs cannot fall, however durable the handle proves to be, thus, financial viability is unlikely to be achieved without intervention in the pricing of hybrid components.

\section{Discussion and Conclusion}
\label{sec:conclusion}

This paper introduced a novel hybrid simulation model combining Discrete-event simulation and Agent-based simulation approaches to represent a circular supply chain network. Conceptualising the supply and recovery chain in this way offers an effective approach to represent the interactions between individual supply chain members and to depict their individual behaviour based on internal policies. This is different to the way circular supply chains in sectors such as apparels, automotive, and consumer durables among others have been modelled in \cite{Amico2024, Huster2022, Lieder2017}, respectively. 

Considering, our model results it is clear that the transition to CE in healthcare supply chains is determined by interconnected financial, environmental, and operational outcomes. These outcomes are influenced not only by product type, but also the CE  strategy (reprocessing of single use, hybrid or fully reusable scissors) adopted, inventory policies (such as the hospital re-order points required to maintain a 100\% service level), and the supporting infrastructure (decontamination and storage capacities along with staff time) required for reprocessing activities. Interestingly, no single CE strategy dominates across all outcomes, highlighting the inherently multi-objective nature of decisions in circular supply chains. 

\subsection{Managerial and Policy Implications}

Across the scenarios analysed, distinct performance patterns emerge. Remanufacturing of SU laparoscopic scissors provides the most immediate financial and environmental benefits, particularly when adoption is prioritised. Fully RU laparoscopic scissors deliver the greatest long-term reductions in CO$_2$ emissions but require extended use cycles to achieve financial viability. Hybrid laparoscopic scissors achieve early environmental benefits but remain economically infeasible under the cost structures assumed here. These findings offer more extensive insights against the predominantly LCA-driven literature on medical device sustainability  \citep{Sassanelli2019, Sassanelli2020}.

The remanufacturing option demonstrates strong alignment between financial and environmental performance. Scenario 1b (prioritised adoption) achieves reductions of approximately 41\% in cost and 55\% in emissions relative to the base scenario, confirming that operational policy plays a crucial role in determining CE outcomes. The additional gains observed due to controlled adoption (Scenario 1a) originate purely from procurement decisions, rather than from product redesign or technological changes.
This finding highlights that CE outcomes are not solely a function of product characteristics, but also of how circular strategies are implemented and managed in the hospital. It reinforces existing findings that EoL strategies can deliver substantial benefits when effectively deployed \cite{Amico2024, Huster2022, Lieder2016}.
However, the feasibility of this options is limited by regulations and product design standards. Current medical device regulations do not support widespread remanufacturing of products certified as single-use, and these products are not typically designed for disassembly or repeated processing. Realising the benefits identified in this study would therefore require upstream design changes and regulatory adaptation, as noted by \cite{Hoveling2024}.

The hybrid laparoscopic CE option highlights a dichotomy between environmental and economic performance. While environmental benefits are achieved —with breakeven occurring after only two use cycles—it remains financially unattractive across all use cycles, which extend to up to 70 cycles.

This finding contradicts the findings of  existing LCA-based studies, particularly \cite{rizan2022environmental}, and can be explained by methodological differences. The simulation model captures supply chain-wide costs, including handling, logistics, and stochastic variability, whereas LCA studies are typically limited at procedure- or product-level analysis. In addition, the use of cost distributions rather than single-point estimates results in higher aggregate cost estimates.

The implication is not that LCA-based findings are incorrect, but that they are limited to their methodological scope. Supply chain-level dynamics can significantly alter economic outcomes, indicating that decisions based solely on LCA cost estimates may lead to suboptimal procurement strategies. This highlights the benefits of adopting simulation approach, as posited also by \cite{Sassanelli2019} that system-level evaluation is necessary for CE decision-making.

Fully reusable laparoscopic scissors offer the greatest environmental benefits, achieving reductions of up to 86\% in emissions at high use cycles, consistent with prior studies \cite{Boberg2022}. However, these benefits are accompanied by delayed financial returns, with breakeven only achieved after approximately 45–50 use cycles.
A key insight is the asymmetry between environmental and financial performance: environmental benefits are realised relatively early (at around three use cycles), while financial viability depends on long-term utilisation. This creates a temporal mismatch that may influence adoption decisions in practice.

The analysis also reveals significant operational implications associated with the adoption of reusable (and hybrid) laparoscopic scissors. These include decisions related to decontamination, handling, storage, and logistics of circular devices. The results show that staff time requirements and processing capacity constraints become important considerations at higher use cycles. Consequently, successful CE implementation requires not only product adoption but also necessary investments in operational infrastructure and workforce capability \citep{Suzanne2020}.

\subsection{Recommendations for stakeholders in the supply chain}

The findings provide actionable insights for key stakeholders within the SMD supply chain.

For hospital procurement managers, CE adoption involves trade-offs between cost, environmental performance, and operational feasibility. The selection of instrument type determines the long-term trajectory of these outcomes, while adoption policies influence the speed at which benefits are realised. The simulation–optimisation results further show that inventory policies must be adapted to account for variability in reprocessing lead times. Failure to do so may result in stockouts, which in healthcare settings translate into cancelled procedures.

In addition, CE adoption introduces new operational requirements, including increased staff handling time and capacity needs for decontamination. These factors must be incorporated into planning decisions, as they represent a shift away from traditional inventory management practices.

For device manufacturers, the results highlight the importance of product design in enabling CE. The analysis shows that use cycles are the most important determinant of cost performance, followed by component cost structure. In the hybrid system, the recurring cost of disposable components remains the primary economic constraint, indicating that design innovations targeting these elements could significantly improve feasibility.

For policymakers, two key recommendations emerge. First,  regulatory changes  enabling remanufacturing of certified single-use devices could unlock significant economic and environmental benefits. Second, targeted financial incentives or procurement policies are required to support the adoption of reusable products, particularly given their high initial costs and delayed financial returns.

\subsection{Limitations}

Of course, the findings presented in this study, need to consider the limitations related to the model developed. The model assumes 100\% instrument collection for remanufacturing, which means that it may be overstating the benefits of Scenario 1. All cost and emissions parameters are held constant, excluding the progressive decarbonisation of the UK electricity grid. Operational parameters, such as handling time,  were expert-informed rather than empirically validated from real world systems as this data is not readily available. Our findings are specific to laparoscopic scissors, which can be adapted to other types of small medical device. Future work could extend the modelling approach presented in this paper to other devices. Furthermore it can incorporate adoption behaviour dynamics, to evaluate the effect of emerging regulatory incentive mechanisms on the long-term viability of each circular strategy.

\section*{Declaration of AI Use}
 
The authors used a generative artificial intelligence tool (Microsoft Copilot) to improve the language, grammar, and readability of this article. The originality and accuracy of all the ideas, interpretations, arguments, including the reported results, data, and
references, remains the full responsibility of the authors.

\section*{Funding}The authors acknowledge that the work presented in this article is part of the “Circular Economy for Small Medical Devices” (ReMed) research project,  funded by the Engineering and Physical Sciences Research Council (EPSRC) (EP/W002566/1). 

\bibliography{interacttfqsample}

@article{Amico2024,
   author = {Clarissa Amico and Roberto Cigolini and Mattia Brambilla},
   doi = {10.1080/09537287.2023.2294301},
   issn = {13665871},
   journal = {Production Planning \& Control},
   publisher = {Taylor \& Francis},
   title = {Transitioning the apparel supply chain to circular economy: a quantitative approach},
   url = {https://www.tandfonline.com/doi/abs/10.1080/09537287.2023.2294301},
   year = {2024}
}

@article{Bassi2021,
   author = {Andrea Marcello Bassi and Marco Bianchi and Marco Guzzetti and Georg Pallaske and Carlos Tapia},
   doi = {10.1016/J.SPC.2020.10.028},
   issn = {2352-5509},
   journal = {Sustainable Production and Consumption},
   month = {7},
   pages = {128-140},
   publisher = {Elsevier},
   title = {Improving the understanding of circular economy potential at territorial level using systems thinking},
   volume = {27},
   url = {https://www.sciencedirect.com/science/article/pii/S2352550920313683},
   year = {2021}
}

@inproceedings{Biller2010,
   author = {Bahar Biller and Canan Gunes},
   editor = {B Johansson and Jain. S and J Montoya-Torres and J Hugan and E Yucesan},
   isbn = {9781424498659},
   booktitle = {2010 Winter Simulation Conference},
   pages = {49-58},
   publisher = {IEEE press},
   title = {Introduction to Simulation Input Modeling},
   year = {2010}
}

@article{Bimpizas-Pinis2022,
   author = {Meletios Bimpizas-Pinis and Tommaso Calzolari and Andrea Genovese},
   doi = {10.1016/j.ijpe.2022.108666},
   issn = {09255273},
   journal = {International Journal of Production Economics},
   month = {8},
   publisher = {Elsevier B.V.},
   title = {Exploring the transition towards circular supply chains through the arcs of integration},
   volume = {250},
   year = {2022}
}

@article{Boberg2022,
   author = {Linn Boberg and Jagdeep Singh and Agneta Montgomery and Peter Bentzer},
   doi = {10.1371/journal.pone.0271601},
   issn = {19326203},
   issue = {7 July},
   journal = {PLoS ONE},
   month = {7},
   pmid = {35839237},
   publisher = {Public Library of Science},
   title = {Environmental impact of single-use, reusable, and mixed trocar systems used for laparoscopic cholecystectomies},
   volume = {17},
   year = {2022}
}

@article{Brailsford2009,
   author = {S. C. Brailsford and P. R. Harper and M. Pitt},
   doi = {10.1057/jos.2009.10},
   issn = {17477786},
   issue = {3},
   journal = {Journal of Simulation},
   pages = {130-140},
   title = {An analysis of the academic literature on simulation and modelling in health care},
   volume = {3},
   year = {2009}
}

@misc{Brailsford2019,
   author = {Sally C. Brailsford and Tillal Eldabi and Martin Kunc and Navonil Mustafee and Andres F. Osorio},
   doi = {10.1016/j.ejor.2018.10.025},
   issn = {03772217},
   issue = {3},
   journal = {European Journal of Operational Research},
   month = {11},
   pages = {721-737},
   publisher = {Elsevier B.V.},
   title = {Hybrid simulation modelling in operational research: A state-of-the-art review},
   volume = {278},
   year = {2019}
}

@inproceedings{Carson2005,
   author = {John S Carson},
   booktitle = {2005 Winter Simulation Conference (WSC)},
   pages = {16-23},
   title = {Introduction to modeling and simulation},
   year = {2005}
}

@article{Charnley2019,
   author = {Fiona Charnley and Divya Tiwari and Windo Hutabarat and Mariale Moreno and Okechukwu Okorie and Ashutosh Tiwari},
   doi = {10.3390/su11123379},
   issue = {12},
   journal = {Sustainability},
   month = {6},
   pages = {3379},
   publisher = {MDPI AG},
   title = {Simulation to Enable a Data-Driven Circular Economy},
   volume = {11},
   year = {2019}
}

@article{Chilmon2020,
   author = {Barbara Chilmon and Nicoleta S. Tipi},
   doi = {10.1016/j.cie.2020.106870},
   issn = {03608352},
   journal = {Computers and Industrial Engineering},
   month = {12},
   publisher = {Elsevier Ltd},
   title = {Modelling and simulation considerations for an end-to-end supply chain system},
   volume = {150},
   year = {2020}
}

@article{Curran2013,
   author = {Mary Ann Curran},
   doi = {10.1016/J.COCHE.2013.02.002},
   issn = {2211-3398},
   issue = {3},
   journal = {Current Opinion in Chemical Engineering},
   month = {8},
   pages = {273-277},
   publisher = {Elsevier},
   title = {Life Cycle Assessment: a review of the methodology and its application to sustainability},
   volume = {2},
   year = {2013}
}

@article{Crdoba-Pachn2025,
   author = {José Rodrigo Córdoba-Pachón and Alberto Paucar-Caceres and Toni Burrowes-Cromwell and Angela Bywater and Ronak Shah and Mark Walker and Kourosh Behzadian},
   doi = {10.1080/01605682.2025.2543445;CTYPE:STRING:JOURNAL},
   issn = {14769360},
   journal = {Journal of the Operational Research Society},
   publisher = {Taylor and Francis Ltd.},
   title = {A critical systems thinking methodology to explore circularity of food waste in a university campus},
   url = {https://www.tandfonline.com/doi/pdf/10.1080/01605682.2025.2543445},
   year = {2025}
}

@article{deOlaeta2023,
   author = {Javier de Olañeta and John Patsavellas and Konstantinos Salonitis},
   doi = {10.1016/J.PROCIR.2023.12.005},
   issn = {2212-8271},
   journal = {Procedia CIRP},
   month = {1},
   pages = {1618-1623},
   publisher = {Elsevier},
   title = {Modelling the operations of a circular economy fashion start-up},
   volume = {120},
   year = {2023}
}

@misc{DepartmentforEnergySecurityandNetZero2024,
   author = {Department for Energy Security and Net Zero},
   title = {Government conversion factors for company reporting of greenhouse gas emissions},
   url = {https://www.gov.uk/government/collections/government-conversion-factors-for-company-reporting},
   year = {2024}
}

@article{Drew2022,
   author = {Jonathan Drew and Sean D. Christie and Daniel Rainham and Chantelle Rizan},
   doi = {10.1016/S2542-5196(22)00257-1/ASSET/AFB69AEF-176B-47AE-A1C8-FAC9CEA10C59/MAIN.ASSETS/GR4.JPG},
   issn = {25425196},
   issue = {12},
   journal = {The Lancet Planetary Health},
   month = {12},
   pages = {e1000-e1012},
   pmid = {36495883},
   publisher = {Elsevier B.V.},
   title = {HealthcareLCA: an open-access living database of health-care environmental impact assessments},
   volume = {6},
   url = {http://www.thelancet.com/article/S2542519622002571/fulltext http://www.thelancet.com/article/S2542519622002571/abstract https://www.thelancet.com/journals/lanplh/article/PIIS2542-5196(22)00257-1/abstract},
   year = {2022}
}

@article{Echefaj2025,
   author = {Khadija Echefaj and Abdelkabir Charkaoui and Anass Cherrafi and Sunil Tiwari and Pankaj Sharma and Charbel Jose Chiappetta Jabbour},
   doi = {10.1080/09537287.2024.2302479},
   issn = {13665871},
   issue = {6},
   journal = {Production Planning and Control},
   pages = {723-747},
   publisher = {Taylor and Francis Ltd.},
   title = {From linear to circular sustainable supply chain network optimisation: towards a conceptual framework},
   volume = {36},
   url = {/doi/pdf/10.1080/09537287.2024.2302479?download=true},
   year = {2025}
}

@article{Fernndez2025,
   author = {María del Valle Fernández and José Manuel Robles and Marta Tolentino and Silvia M. Andrade},
   doi = {10.1080/23311975.2025.2499668},
   issn = {23311975},
   issue = {1},
   journal = {Cogent Business and Management},
   publisher = {Cogent OA},
   title = {Analysis of the degree of implementation of the circular economy in Europe and Spain},
   volume = {12},
   year = {2025}
}

@article{Gajanayake2025,
   author = {Akvan Gajanayake and Usha Iyer-Raniga},
   doi = {10.1016/J.ECOLECON.2024.108395},
   issn = {0921-8009},
   journal = {Ecological Economics},
   month = {1},
   pages = {108395},
   publisher = {Elsevier},
   title = {If there is waste, there is a system: Understanding Victoria's circular economy transition from a systems thinking perspective},
   volume = {227},
   url = {https://www.sciencedirect.com/science/article/pii/S0921800924002921},
   year = {2025}
}

@inbook{Goldsman2015,
   author = {David Goldsman and Paul Goldsman},
   doi = {10.1007/978-1-4471-5634-5_10},
   editor = {M Loper},
   isbn = {978-1-4471-5634-5},
   issn = {2195-2825},
   booktitle = {Modeling and Simulation in the Systems Engineering Life Cycle},
   pages = {103-109},
   publisher = {Springer, London},
   title = {Discrete-Event Simulation},
   url = {https://link.springer.com/chapter/10.1007/978-1-4471-5634-5_10},
   year = {2015}
}

@inbook{Guinee2017,
   author = {Jeroen Guinee and Reinout Heijungs},
   editor = {Yann Bouchery and Charles J Corbett and Jan C Fransoo and Tarkan Tan},
   booktitle = {Sustainable Supply Chains},
   pages = {15-42},
   publisher = {Springer},
   title = {Introduction to Life Cycle Assessment},
   volume = {4},
   year = {2017}
}

@inproceedings{Guzzo2019,
   author = {D Guzzo and E Jamsin and R Balkenende and J.M.H Costa},
   doi = {10.14279/depositonce-9253},
   isbn = {978-3-7983-3125-9},
   booktitle = {PLATE},
   pages = {309-315},
   title = {The Use of System Dynamics to Verify Long-term Behaviour and Impacts of Circular Business Models: a Sharing Platform in Healthcare},
   url = {https://doi.org/10.14279/},
   year = {2019}
}

@article{Haupt2017,
   author = {Melanie Haupt and Mischa Zschokke},
   doi = {10.1007/S11367-017-1267-1/FIGURES/1},
   issn = {16147502},
   issue = {5},
   journal = {International Journal of Life Cycle Assessment},
   month = {5},
   pages = {832-837},
   publisher = {Springer Verlag},
   title = {How can LCA support the circular economy?—63rd discussion forum on life cycle assessment, Zurich, Switzerland, November 30, 2016},
   volume = {22},
   url = {https://link.springer.com/article/10.1007/s11367-017-1267-1},
   year = {2017}
}

@article{Hoveling2024,
   author = {Tamara Hoveling and Anne Svindland Nijdam and Marlou Monincx and Jeremy Faludi and Conny Bakker},
   doi = {10.1016/J.RESCONREC.2024.107719},
   issn = {0921-3449},
   journal = {Resources, Conservation and Recycling},
   month = {9},
   pages = {107719},
   publisher = {Elsevier},
   title = {Circular economy for medical devices: Barriers, opportunities and best practices from a design perspective},
   volume = {208},
   year = {2024}
}

@article{Huster2022,
   author = {Sandra Huster and Simon Glöser-Chahoud and Sonja Rosenberg and Frank Schultmann},
   doi = {10.1016/J.JCLEPRO.2022.132225},
   issn = {0959-6526},
   journal = {Journal of Cleaner Production},
   month = {8},
   pages = {132225},
   publisher = {Elsevier},
   title = {A simulation model for assessing the potential of remanufacturing electric vehicle batteries as spare parts},
   volume = {363},
   year = {2022}
}

@article{Katsaliaki2011,
   author = {K. Katsaliaki and N. Mustafee},
   doi = {10.1057/jors.2010.20},
   issn = {14769360},
   issue = {8},
   journal = {Journal of the Operational Research Society},
   pages = {1431-1451},
   publisher = {Palgrave Macmillan Ltd.},
   title = {Applications of simulation within the healthcare context},
   volume = {62},
   year = {2011}
}

@article{Keselman1977,
   author = {H. J. Keselman and Joanne C. Rogan},
   doi = {10.1037/0033-2909.84.5.1050},
   issn = {00332909},
   issue = {5},
   journal = {Psychological Bulletin},
   month = {9},
   pages = {1050-1056},
   title = {The Tukey multiple comparison test: 1953-1976},
   volume = {84},
   year = {1977}
}

@article{Konstantaras2021,
   author = {Ioannis Konstantaras and Konstantina Skouri and Lakdere Benkherouf},
   doi = {10.1016/J.IJPE.2021.108185},
   issn = {0925-5273},
   journal = {International Journal of Production Economics},
   month = {9},
   pages = {108185},
   publisher = {Elsevier},
   title = {Optimizing inventory decisions for a closed–loop supply chain model under a carbon tax regulatory mechanism},
   volume = {239},
   url = {https://www.sciencedirect.com/science/article/abs/pii/S0925527321001614?casa_token=h-XkzAeU8wcAAAAA:ehn2-AB7_ndfI8OUe3ZJHXTYktDjvotM1Q5PZZj6Rk6Y6GLJT2E2avSG-Nu11inM_1NNfKMc},
   year = {2021}
}

@article{Kramer2023,
   author = {Kathrin Julia Kramer and Matthias Schmidt},
   doi = {10.1007/978-3-031-28839-5_10/FIGURES/4},
   isbn = {9783031288388},
   issn = {21954364},
   journal = {Lecture Notes in Mechanical Engineering},
   pages = {85-93},
   publisher = {Springer Science and Business Media Deutschland GmbH},
   title = {Circular Supply Chain Management in the Wind Energy Industry – A Systematic Literature Review},
   url = {https://link.springer.com/chapter/10.1007/978-3-031-28839-5_10},
   year = {2023}
}

@article{rizan2022environmental,
  title={Environmental impact and life cycle financial cost of hybrid (reusable/single-use) instruments versus single-use equivalents in laparoscopic cholecystectomy},
  author={Rizan, Chantelle and Bhutta, Mahmood F},
  journal={Surgical Endoscopy},
  volume={36},
  number={6},
  pages={4067--4078},
  year={2022},
  publisher={Springer}
}

@article{Labib2024,
   author = {P. L. Labib and B. S.M. Ford and M. Winfield and W. J. Douie and A. Kanwar and G. Sanders},
   doi = {10.1308/rcsann.2023.0015},
   issn = {14787083},
   issue = {2},
   journal = {Annals of the Royal College of Surgeons of England},
   month = {2},
   pages = {167-172},
   pmid = {37051744},
   publisher = {Royal College of Surgeons of England},
   title = {Revising a laparoscopic appendicectomy set to reduce reliance on disposable surgical instruments: supporting the transition to sustainable surgical practice},
   volume = {106},
   year = {2024}
}

@article{Leiden2020,
   author = {Alexander Leiden and Felipe Cerdas and David Noriega and Jörg Beyerlein and Christoph Herrmann},
   doi = {10.1016/J.RESCONREC.2020.104704},
   issn = {0921-3449},
   journal = {Resources, Conservation and Recycling},
   month = {5},
   pages = {104704},
   publisher = {Elsevier},
   title = {Life cycle assessment of a disposable and a reusable surgery instrument set for spinal fusion surgeries},
   volume = {156},
   year = {2020}
}

@article{Lewandowski2016,
   author = {Mateusz Lewandowski},
   doi = {10.3390/SU8010043},
   issn = {2071-1050},
   issue = {1},
   journal = {Sustainability 2016, Vol. 8, Page 43},
   month = {1},
   pages = {43},
   publisher = {Multidisciplinary Digital Publishing Institute},
   title = {Designing the Business Models for Circular Economy—Towards the Conceptual Framework},
   volume = {8},
   url = {https://www.mdpi.com/2071-1050/8/1/43/htm https://www.mdpi.com/2071-1050/8/1/43},
   year = {2016}
}

@misc{Nurse,
  author = {NHS Employers},
  title = {Pay scales for 2023/24},
  year = {2023},
  note = {Last accessed 16 September 2024},
  url = {https://www.nhsemployers.org/articles/pay-scales-202324-archived}
}

@misc{sea-distance,
  author = {Online},
  title = {sea-distance},
  year = {2023},
  note = {Last accessed 16 September 2024},
  url = {https://sea-distances.org/}
}

@misc{Lieder2016,
   author = {Michael Lieder and Amir Rashid},
   doi = {10.1016/j.jclepro.2015.12.042},
   issn = {09596526},
   journal = {Journal of Cleaner Production},
   month = {3},
   pages = {36-51},
   publisher = {Elsevier Ltd},
   title = {Towards circular economy implementation: A comprehensive review in context of manufacturing industry},
   volume = {115},
   year = {2016}
}

@article{Lieder2017,
   author = {Michael Lieder and Farazee M.A. Asif and Amir Rashid and Aleš Mihelič and Simon Kotnik},
   doi = {10.1007/s00170-017-0610-9},
   issn = {14333015},
   issue = {5-8},
   journal = {International Journal of Advanced Manufacturing Technology},
   month = {11},
   pages = {1953-1970},
   publisher = {Springer London},
   title = {Towards circular economy implementation in manufacturing systems using a multi-method simulation approach to link design and business strategy},
   volume = {93},
   year = {2017}
}

@article{Linder2017,
   author = {Marcus Linder and Mats Williander},
   doi = {10.1002/BSE.1906},
   issn = {10990836},
   issue = {2},
   journal = {Business Strategy and the Environment},
   month = {2},
   pages = {182-196},
   publisher = {John Wiley and Sons Ltd},
   title = {Circular Business Model Innovation: Inherent Uncertainties},
   volume = {26},
   year = {2017}
}

@article{Macal2009,
   author = {Charles M. Macal and Michael J. North},
   doi = {10.1109/WSC.2009.5429318},
   isbn = {9781424457700},
   issn = {08917736},
   journal = {Proceedings - Winter Simulation Conference},
   pages = {86-98},
   title = {Agent-based modeling and simulation},
   year = {2009}
}

@misc{Moon2017,
   author = {Young B. Moon},
   doi = {10.1080/19397038.2016.1220990},
   issn = {19397046},
   issue = {1},
   journal = {International Journal of Sustainable Engineering},
   month = {1},
   pages = {2-19},
   publisher = {Taylor and Francis Ltd.},
   title = {Simulation modelling for sustainability: a review of the literature},
   volume = {10},
   year = {2017}
}

@article{Pea2021,
   author = {Claudia Peña and Bárbara Civit and Alejandro Gallego-Schmid and Angela Druckman and Armando Caldeira-Pires and Bo Weidema and Eric Mieras and Feng Wang and Jim Fava and Llorenç Milà i. Canals and Mauro Cordella and Peter Arbuckle and Sonia Valdivia and Sophie Fallaha and Wladmir Motta},
   doi = {10.1007/S11367-020-01856-Z/METRICS},
   issn = {16147502},
   issue = {2},
   journal = {International Journal of Life Cycle Assessment},
   month = {2},
   pages = {215-220},
   publisher = {Springer Science and Business Media Deutschland GmbH},
   title = {Using life cycle assessment to achieve a circular economy},
   volume = {26},
   url = {https://link.springer.com/article/10.1007/s11367-020-01856-z},
   year = {2021}
}

@article{Randersed1997,
   author = {J. Randers (ed)},
   doi = {10.1057/palgrave.jors.2600456},
   issn = {14769360},
   issue = {11},
   journal = {Journal of the Operational Research Society},
   pages = {1144-1145},
   title = {Elements of the system dynamics method},
   volume = {48},
   year = {1997}
}

@inproceedings{sargent2010verification,
  title={Verification and validation of simulation models},
  author={Sargent, Robert G},
  booktitle={Proceedings of the 2010 winter simulation conference},
  pages={166--183},
  year={2010},
  organization={IEEE}
}

@INPROCEEDINGS{shoaibesc,
  author={Shoaib, Mohd and Tako, Antuela and Fayad, Ramzi and Vargas-Palacios, Armando},
  booktitle={2024 Winter Simulation Conference (WSC)}, 
  title={Towards using Simulation to Evaluate the Circular Economy of Small Medical Devices}, 
  year={2024},
  volume={},
  number={},
  pages={1059-1070},
  doi={10.1109/WSC63780.2024.10838979}}

@book{robinson2025simulation,
  title={Simulation: the practice of model development and use},
  author={Robinson, Stewart},
  year={2025},
  publisher={Bloomsbury Publishing}
}

@inproceedings{balci1997verification,
  title={Verification validation and accreditation of simulation models},
  author={Balci, Osman},
  booktitle={Proceedings of the 29th conference on Winter simulation},
  pages={135--141},
  year={1997}
}

@book{Robinson1994,
   author = {S Robinson},
   city = {Maidenhead, UK},
   publisher = {McGraw-Hill},
   title = {Successful Simulation: A Practical Approach to Simulation Projects},
   year = {1994}
}

@misc{Sassanelli2019,
   author = {Claudio Sassanelli and Paolo Rosa and Roberto Rocca and Sergio Terzi},
   doi = {10.1016/j.jclepro.2019.05.019},
   issn = {09596526},
   journal = {Journal of Cleaner Production},
   month = {8},
   pages = {440-453},
   publisher = {Elsevier Ltd},
   title = {Circular economy performance assessment methods: A systematic literature review},
   volume = {229},
   year = {2019}
}

@inproceedings{Sassanelli2020,
   author = {Claudio Sassanelli and Paolo Rosa and Sergio Terzi},
   doi = {10.5220/0009989300530059},
   issn = {21849285},
   booktitle = {IN4PL - Proceedings of the International Conference on Innovative Intelligent Industrial Production and Logistics},
   pages = {53-59},
   title = {Circular economy-oriented simulation: A literature review grounded on empirical cases},
   year = {2020}
}

@article{Suzanne2020,
   author = {Elodie Suzanne and Nabil Absi and Valeria Borodin},
   doi = {10.1016/J.EJOR.2020.04.043},
   issn = {0377-2217},
   issue = {1},
   journal = {European Journal of Operational Research},
   month = {11},
   pages = {168-190},
   publisher = {North-Holland},
   title = {Towards circular economy in production planning: Challenges and opportunities},
   volume = {287},
   url = {https://www.sciencedirect.com/science/article/abs/pii/S0377221720303969?via%3Dihub},
   year = {2020}
}

@article{Tako2012,
   author = {Antuela A. Tako and Stewart Robinson},
   doi = {10.1016/j.dss.2011.11.015},
   issn = {01679236},
   issue = {4},
   journal = {Decision Support Systems},
   month = {3},
   pages = {802-815},
   title = {The application of discrete event simulation and system dynamics in the logistics and supply chain context},
   volume = {52},
   year = {2012}
}

@article{Taylor2009,
   author = {S. J.E. Taylor and T. Eldabi and G. Riley and R. J. Paul and M. Pidd},
   doi = {10.1057/JORS.2008.196},
   issn = {14769360},
   issue = {SUPPL. 1},
   journal = {Journal of the Operational Research Society},
   pages = {69-82},
   publisher = {Taylor \& Francis},
   title = {Simulation modelling is 50! Do we need a reality check?},
   volume = {60},
   url = {https://www.tandfonline.com/doi/abs/10.1057/jors.2008.196},
   year = {2009}
}

@article{Walzberg2020,
 
   author = {Julien Walzberg and Geoffrey Lonca and Rebecca J. Hanes and Annika L. Eberle and Alberta Carpenter and Garvin A. Heath},
   doi = {10.3389/frsus.2020.620047},
   issn = {26734524},
   journal = {Frontiers in Sustainability},
   publisher = {Frontiers Media S.A.},
   title = {Do We Need a New Sustainability Assessment Method for the Circular Economy? A Critical Literature Review},
   volume = {1},
   year = {2020}
}

@article{Walzberg2021,
   
   author = {Julien Walzberg and Alberta Carpenter and Garvin A. Heath},
   doi = {10.1038/s41560-021-00888-5},
   issn = {20587546},
   issue = {9},
   journal = {Nature Energy},
   month = {9},
   pages = {913-924},
   publisher = {Nature Research},
   title = {Role of the social factors in success of solar photovoltaic reuse and recycle programmes},
   volume = {6},
   year = {2021}
}

@article{YahyapourGanji2025,
 
   author = {Vahid Yahyapour Ganji and Ehsan Hozan and Parisa Babolhavaeji and Amir Reza Tajally and Mohssen Ghanavati-Nejad},
   doi = {10.1016/j.seps.2025.102294},
   issn = {00380121},
   journal = {Socio-Economic Planning Sciences},
   month = {10},
   publisher = {Elsevier Ltd},
   title = {A robust design of a circular supply chain network based on the resilience and responsiveness dimensions: A data-driven model},
   volume = {101},
   year = {2025}
}

\appendix

\section{Literature Summary}
\label{app:lit-summary}

\begin{uselscape}

\begin{table}[p]
\centering
\caption{Summary of simulation-based studies on circular economy
         transitions across sectors.}
\label{tab:litreview}
\footnotesize
\begin{threeparttable}

  \begin{tabular}{
      L{2.2cm}   
      L{5cm}   
      L{2.3cm}  
      L{2.3cm}   
      L{2.6cm}   
      L{5.3cm}   
    }

  \toprule
  \textbf{Publication}
    & \textbf{Problem studied}
    & \textbf{Methodology}
    & \textbf{Sector / industry}
    & \textbf{Product / system}
    & \textbf{Key findings} \\
  \midrule

  Ameli et al. (2019)
    & Optimising EoL strategy selection for product design
    & DES \& multi-objective optimisation
    & Electronics
    & Printed circuit boards
    & Integrates sustainability metrics in design-stage decisions \\[6pt]

  Amico et al. (2024)
    & Assessing effects of reuse, remanufacture, and recycling
      strategies on apparel supply chain performance
    & DES
    & Apparel supply chain
    & Jackets
    & Reuse yields highest profit; joint reuse--recycle improves
      environmental and economic performance \\[6pt]

  Castiglione et al. (2025)
    & Scheduling-level challenges when integrating disassembly
      and remanufacturing
    & DES
    & Manufacturing
    & Generic products
    & Frequent rescheduling needed due to variability in returns \\[6pt]

  Charnley et al. (2019)
    & How return quality affects remanufacturing process times
      in an EVB facility
    & DES
    & Automotive
    & EV batteries
    & Variable quality significantly affects system throughput and
      processing times \\[6pt]

  de Ola\~neta et al. (2023)
    & Operational performance of a subscription-based CE business
      model for fashion rentals
    & DES
    & Apparel
    & Clothing
    & Inventory and cash flow optimised via tailored subscription
      parameters \\[6pt]

  Fani et al. (2025)
    & Consumer behaviour and CE compliance for medication take-back
    & ABS
    & Healthcare
    & Pharmaceutical
    & Behavioural interventions significantly improve collection
      outcomes \\[6pt]

  Ganji et al. (2025)
    & Design of a resilient, circular, and responsive supply chain
      network
    & Mathematical model
    & Healthcare
    & Haemodialysis devices
    & Optimal facility locations identified under mixed uncertainty \\[6pt]

  Guzzo et al. (2020)
    & Sharing-based business model for hospital devices and its
      operational implications
    & SD
    & Healthcare
    & Hospital devices
    & Sharing reduces new product acquisition and unmet demand \\[6pt]

  Hanes et al. (2021)
    & Evaluating circularity scenarios for wind turbines using
      lifecycle and system-wide modelling
    & DES \& LCA
    & Energy
    & Wind turbine blades
    & Circular strategies markedly reduce lifecycle emissions and
      improve resource circularity \\[6pt]

  Huster et al. (2022)
    & Impact of remanufacturing EV batteries on demand for new
      batteries
    & DES
    & Automotive
    & Electric vehicle (EV) batteries
    & Remanufacturing reduces new battery demand under multiple
      lifetime scenarios \\[6pt]

  Lieder et al. (2017)
    & Combined CE strategies and business models for washing
      machines
    & ABS \& DES
    & Consumer durables
    & Washing machines
    & Leasing / pay-per-use models reduce emissions and lifecycle
      costs \\[6pt]

  Roci et al. (2022)
    & Multi-method simulation for CE transitions in manufacturing
    & ABM, DES, \& SD
    & Manufacturing
    & Washing machines
    & Combining SD and DES improves representation of CE feedback
      effects \\[6pt]

  Walzberg et al. (2021)
    & Linking CE strategies and system-wide environmental
      performance
    & ABM
    & Energy
    & Photovoltaic modules
    & Consumer behaviour change essential to promote reuse;
      recycling should complement reuse \\[4pt]

  \bottomrule
  \end{tabular}

  \begin{tablenotes}[flushleft]\footnotesize
    \item EoL: end of life; DES: discrete-event simulation;
          ABS/ABM: agent-based simulation/modelling; SD: system dynamics;
          LCA: life cycle assessment; CE: circular economy;
          EVB: electric vehicle battery.
  \end{tablenotes}

\end{threeparttable}
\end{table}
\end{uselscape}

\section{Methodology and Material}

\subsection{Supply Chain Member Operations and Decision Logic}
\label{app:methodology_member}

This section details the operational logic implemented at each supply chain member, complementing the model overview presented in Section \ref{sec:methodology-overview} and Figure \ref{fig:conceptual-diagram} .

\subsubsection*{Hospital Agent}
 
Demand is generated by patients arrival for surgery.  The following functions are then executed.
 
\begin{itemize}[leftmargin=*, label={o}]
 
  \item \textit{Distribution Centre (DC) selection.}
        A hospital agent selects only one DC per product category.
        Where more than one DC stocks the same product, selection is
        based on the shortest distance criterion.
 
  \item \textit{Inventory management and ordering.}
        Hospitals hold stock equivalent to 4--6~days of demand. A Min-Max $(s,\,S)$ policy with continuous review is
        implemented.
        The maximum inventory level~$S$ is sampled from a
        $\mathrm{Uniform}(4,6)$ distribution and multiplied by the
        expected number of daily procedures.
        The reorder point~$s$ is calculated as
        
        \begin{equation}
          s \;=\; \bar{d}\cdot\overline{LT} + SS,
          \label{eq:reorder}
        \end{equation}
        
        where $\bar{d}$ is average demand per day and
        $\overline{LT}$ is average lead time.
        Safety stock is computed as
        
        \begin{equation}
          SS \;=\; z\,\sqrt{\sigma_{d}^{2}\cdot\overline{LT}
                            \;+\;
                            \sigma_{LT}^{2}\cdot\bar{d}^{2}},
          \label{eq:ss}
        \end{equation}
        
        where $z = 3.5$, corresponding to a 99.9\% service level
        standard in healthcare supply chains, and $\sigma_{d}$ and
        $\sigma_{LT}$ are the standard deviations of demand and lead
        time, respectively.
        For low-demand hospitals with fewer than five procedures per
        day, the reorder point is fixed at five instruments and the
        maximum inventory capped at ten.
        When stock on hand falls below~$s$, a new order is placed for
        the quantity equal to the difference between current stock
        and~$S$.
 
  \item \textit{Management of used SMDs.}
        Based on the instrument type, used SMDs are handled
        appropriately. For example, a RU instrument is disassembled, cleaned, packed, and stored, while a SU instrument is stored locally to be incinerated at a later stage.
 
\end{itemize}

\subsubsection*{Distribution Centre (DC)}
 
The primary functions of the DC agent are manufacturer selection, stock monitoring, ordering, and hospital order fulfilment. It is assumed that this is fullfilled via truck transport.
 
\begin{itemize}[leftmargin=*, label={o}]
 
  \item \textit{Manufacturer selection.}
        A DC agent selects only one manufacturer per product.
        Where more than one facility produces the same product, the
        nearest is selected.
 
  \item \textit{Inventory management and ordering.}
        A continuous review Min-Max $(s,\,S)$ policy is implemented.
        Safety stock is calculated as
        $SS = z\,\sigma_{LT}\,\bar{d}$,
        with demand treated as constant but set conservatively higher
        to avoid stockouts, given that the downstream demand
        distribution is not known \textit{a priori}.
        The maximum inventory level is set at $S = 2s$.
        When stock falls below the reorder point, an order is placed
        for the difference between current stock and~$S$.
 
  \item \textit{Order management and transportation.}
        Hospital orders are fulfilled on a first-come first-served
        (FCFS) basis using truck agents.
 
\end{itemize}

\subsubsection*{Manufacturer}
 
The manufacturer agent is responsible for the following:
 
\begin{itemize}[leftmargin=*, label={o}]
 
  \item \textit{Production.}
        The manufacturer produces SU, RU, and hybrid SMDs under a
        make-to-stock strategy.
 
  \item \textit{Inventory management.}
        Stock levels are monitored continuously.
        Production is initiated when stock reaches a prespecified
        minimum threshold.
 
  \item \textit{Order management.}
        Orders are fulfilled on a FCFS basis.
 
\end{itemize}

\subsubsection*{Reprocessor}
 
The reprocessor forms part of the recovery chain and is responsible for remanufacturing used SU  laparascopic scissors, which are
collected from hospitals. Its main tasks are as follows.
 
\begin{itemize}[leftmargin=*, label={o}]
 
  \item \textit{Product collection and quality assessment.}
        Used SMDs are collected, cleaned, decontaminated, and assessed
        to determine whether the instrument has reached its maximum
        allowable use cycles.
 
  \item \textit{Remanufacturing.}
        Used SMDs are processed to a new-like condition and stored for
        dispatch to the DC or hospitals directly.
 
  \item \textit{Order management.}
        Orders are fulfilled on a FIFO basis.
 
\end{itemize}

\subsection{Additional Model Assumptions}
\label{app:assumptions}

\begin{itemize}

    \item The supply chain starts with the manufacturer level; the manufacturer is assumed to hold all necessary materials and components, and all production is conducted at a single facility. Manufacturing emissions are independent of the manufacturer's  location.
 
  \item All cost parameters are assumed to remain constant over the simulation period. Inflation, changes in NHS pay scales, and future instrument pricing are not modelled.
  
  \item One laparoscopic scissor is required per  surgery, and all
  procedures require laparoscopic scissors.
  
  \item For material volume calculations, instrument materials are aggregated into metal and plastic categories; specific alloys and polymer grades are not distinguished.
 
  \item Packaging is excluded from all calculations.
 
  \item For ship transport, only one-way emissions for carrying the devices from Chine to the UK are included.
 
  \item The repair centre is assumed to be at a fixed distance of
  40~km from all hospitals.
\end{itemize}

\subsection{Model Input Data}
\label{app:inputdata}

Table~\ref{tab:independent} lists laparoscopic scissors-independent parameters
common across all scenarios, including nurse handling and disposal
costs, transport emission factors, order lead times, and 
failure probabilities.
Table~\ref{tab:hybrid} lists cost and emission parameters specific to
hybrid scissors covering both the reusable handle and the
disposable blade across manufacturing, repair, and waste disposal
stages.
Table~\ref{tab:ru} provides equivalent parameters for fully RU
laparoscopic scissors.
Table~\ref{tab:su} lists parameters for SU scissors.

\textit{Financial parameters}
To reflect the diverse market of laparoscopic scissors available to the
NHS, cost variability was modelled through consultation with industry
experts.
Laparoscopic scissors were segmented into three tiers: low (10\% of market),
medium (80\%), and high (10\%), each assigned a cost range represented
by a Uniform distribution (Tables~\ref{tab:hybrid}--\ref{tab:ru}).
Direct costs such as repair and decontamination were sourced from the
literature or industry experts and modelled as point estimates.

\textit{Environmental parameters}
Environmental indicators are based on CO$_2$-eq emissions.
Lifecycle emissions, including those from manufacturing, decontamination, and
disposal, were derived from LCA studies \citep{rizan2022environmental, Labib2024}
and used as mode values for Triangular distributions.
Transport emissions were calculated using data from the UK Department
for Energy Security and Net Zero (2024).
Road transport variability was modelled using a Uniform distribution
reflecting emissions between a fully laden
(140.87~gCO$_2$/tonne/km) and 50\% laden
(250.73~gCO$_2$/tonne/km) heavy goods vehicle.
Decontamination emissions were derived from \cite{rizan2022environmental}, based
on decontamination of an average set of approximately twenty
instruments.

\textit{Operational parameters}
Operational parameters are less readily available in the published
literature, as most hospitals currently use SU laparoscopic scissors and
first-hand data on reusable and hybrid laparoscopic scissors handling are largely
unavailable.
These parameters were therefore specified through structured
discussions with industry experts.
Handling time, representing overall staff time for assembly,
disassembly, and internal logistics is modelled as a Uniform
distribution (Table~\ref{tab:independent}), consistent with the range
provided by industry experts.
SU laparoscopic scissors reliability is captured through a fixed failure
probability of 0.1 per use cycle.
For hybrid and RU instruments, repair probability follows a
$\mathrm{Uniform}(0.01,\,0.2)$ distribution
(Tables~\ref{tab:hybrid}--\ref{tab:ru}), reflecting experts' input
that laparoscopic scissors designed for repeated use are more prone to wear than
SU devices.
For fully RU laparoscopic scissors, repair events are further classified as minor
(probability 0.8) or major (probability 0.2), where major repairs may
involve part replacement and minor repairs tasks such as sharpening.

\begin{table}[htbp]
\centering
\caption{Laparoscopic scissors-independent data and input parameter distributions.}
\label{tab:independent}
\small
\begin{threeparttable}

  \begin{tabular}{p{4.2cm} p{3cm} p{6cm} }
  \toprule
  {Parameter} & {Distribution / Value} & Description \& Source \\
  \midrule

  \multicolumn{3}{l}{\textit{Costs in £\,GBP\tnote{a}}} \\[2pt]
  Handling cost for nurse (per hour)
    & $\mathrm{Uniform}(26.06,\;42.74)$
    & 
    (Source: \cite{Nurse})\\[6pt]
  Repair (courier) transport (per instrument)
    & £\,1.00
    & £20 per laparoscopic scissors set;
      assuming 15--20 instruments per set (Source: Industry informed) \\[6pt]
      
  Decontamination (per instrument)
    & £\,1.05
    &  Based on decontamination
      of a laparoscopic scissors set (Source: \cite{rizan2022environmental}) \\[6pt]

  Disposal (per kg)
    & $\mathrm{Gamma}(25,\;0.02469)$\newline
      $0.617 \pm 20\%$
    & Mean $=$ £\,0.617/kg.
      (Source: \cite{rizan2022environmental} \\[6pt]

  \midrule

  \multicolumn{3}{l}{\textit{Emissions in gCO$_2$-eq}} \\[2pt]

  Transport emissions --- diesel HGV\tnote{b}
  (7.5--17 tonnes)
    & $\mathrm{Uniform}(140.87,\;250.73)$
      gCO$_2$/tonne/km
    & 140.87 for 100\% laden;
      250.73 for 50\% laden (Source: \cite{DepartmentforEnergySecurityandNetZero2024}) \\[6pt]

  Freight emissions by cargo ship
    & $13.21\;\mathrm{gCO_2} \times
      \text{total tonne-km}$
    & Average general cargo (Source: \cite{DepartmentforEnergySecurityandNetZero2024}) \\[6pt]

  \midrule

  \multicolumn{3}{l}{\textit{Miscellaneous}} \\[2pt]

  Order lead time from manufacturer
  to DC\tnote{c}
    & $\mathrm{Uniform}(10,\;20)$ days
    & (Source: Industry informed) \\[6pt]

  Handling time for hybrid \& reusable
  laparoscopic scissors (minutes per instrument)
    & $\mathrm{Uniform}(5,\;20)$
    & (Source: Industry informed) \\[6pt]

  Laparoscopic scissors failure probability
    & 0.10
    & Assumption \\[6pt]

  Patient arrivals per day across
  nine hospitals
    & 0.11,\ 7.76,\ 0.78,\ 0.28,\
      6.90,\ 1.60,\ 3.21,\ 7.76,\ 10.00
    & (Source: Freedom of Information request) \\[2pt]

  \bottomrule
  \end{tabular}

  \begin{tablenotes}[flushleft]\footnotesize
    \item[a] Great British Pound (GBP).
    \item[b] Heavy goods vehicle (HGV).
    \item[c] Distribution centre (DC).
  \end{tablenotes}

\end{threeparttable}
\end{table}

\begin{table}[htbp]
\centering
\caption{Hybrid laparoscopic scissors-specific data and input parameter distributions. \textit{Note: For emission parameters, the first line gives the statistical distribution used in the model (e.g., Triangular(min, mode, max)); the second line gives the source value and the percentage spread applied symmetrically around it to derive the distribution bounds}}
\label{tab:hybrid}
\small
\begin{threeparttable}

  \begin{tabular}{
      p{3cm}
      p{4.5cm}
      p{6.5cm}
    }

  \toprule
  {Parameter}
    & {Distribution / Value}
    & {Description \& Source} \\
  \midrule

  \multicolumn{3}{l}{\textit{Costs in £\,GBP}} \\[2pt]

  Reusable component (new)
    & P$(X<
    0.1)$: Uniform(125, 150) \newline 
      P$(0.1\leq X<0.9)$ = Uniform(250, 300) \newline 
      P$(X>0.9)$ = Uniform(350, 400)
    
    & 10\% of the laparoscopic scissors purchased fall in the low price category (between £125 to £150);
    80\% of the laparoscopic scissors lie in the medium price category (between £250 to £300); and
    remaining 10\% in the high cost category (between £350 to £400).
    (Source: Industry informed) \\[6pt]

  Disposable component (new)
    & P$(X<0.1)$: Uniform(16, 20) \newline 
      P$(0.1\leq X < 0.9)$ = Uniform(23, 27) \newline 
      P$(X > 0.9)$ = Uniform(30, 34)

    & 10\% of the laparoscopic scissors purchased fall in the low price category (between £16 to £20);
    80\% of the laparoscopic scissors lie in the medium price category (between £23 to £27); and
    remaining 10\% in the high cost category (between £30 to £34).
    (Source: Industry informed) \\[6pt]

  Repair
    & £30
    & (Source: Industry informed) \\[6pt]

  \midrule

  \multicolumn{3}{l}{\textit{Emissions in gCO$_2$-eq}} \\[2pt]

  Decontamination
    & $\mathrm{Triangular}(63.2,\;79,\;94.8)$\newline
      $79 \pm 15.8\ (20\%)$\tnote{a}
    & Mean $= 79$. (Source: \cite{rizan2022environmental}) \\[6pt]

  Waste disposal (disposable component)
    & $\mathrm{Triangular}(51.2,\;64,\;76.8)$\newline
      $64 \pm 12.8\ (20\%)$\tnote{a}
    & Mean $= 64$. (Source: \cite{rizan2022environmental}) \\[6pt]

  Waste disposal (reusable component)
    & $\mathrm{Triangular}(148,\;185,\;222)$\newline
      $185 \pm 37\ (20\%)$\tnote{a}
    & Mean $= 185$. (Source: \cite{rizan2022environmental}) \\[6pt]

  Manufacturing (disposable component)
    & $\mathrm{Triangular}(185.6,\;232,\;278.4)$\newline
      $232 \pm 46.4\ (20\%)$\tnote{a}
    & Mean $= 232$. (Source: \cite{rizan2022environmental}) \\[6pt]

  Manufacturing (reusable component)
    & $\mathrm{Triangular}(508,\;635,\;762)$\newline
      $635 \pm 127\ (20\%)$\tnote{a}
    & Mean $= 635$. (Source: \cite{rizan2022environmental}) \\[6pt]

  Repair (reusable component)
    & $\mathrm{Triangular}(50.8,\;63.5,\;76.2)$\newline
      $63.5 \pm 12.7\ (20\%)$
    & 10\% of manufacturing
      emissions. (Source: Assumption) \\[6pt]

  \midrule

  \multicolumn{3}{l}{\textit{Miscellaneous}} \\[2pt]

  Reusable component weight
    & 73\,g (Metal: 20.87\,g;
      PPS\tnote{b}: 51.9\,g)
    & (Source: \cite{rizan2022environmental}) \\[6pt]

  Disposable component weight
    & 25.14\,g (Metal: 25\,g)
    & (Source: \cite{rizan2022environmental}) \\[6pt]

  Probability an instrument requires
  repair, $P(R)$
    & $\mathrm{Uniform}(0.01,\;0.2)$
    & (Source: Industry informed) \\[2pt]

  \bottomrule
  \end{tabular}

  \begin{tablenotes}[flushleft]\footnotesize
    \item[a] Modelled via a triangular distribution centred at the mean (mode) with a 20\% variation.
    \item[b] Polyphenylene sulphide (PPS).

  \end{tablenotes}

\end{threeparttable}
\end{table}

\begin{table}[htbp]
\centering
\caption{Fully reusable laparoscopic scissors-specific data and input parameter
         distributions. \textit{Note: For emission parameters, the first line gives the statistical distribution used in the model (e.g., Triangular(min, mode, max)); the second line gives the source value and the percentage spread applied symmetrically around it to derive the distribution bounds}}
\label{tab:ru}
\small
\begin{threeparttable}

  \begin{tabular}{
      p{3.5cm} p{5.1cm} p{5cm}
    }

  \toprule
  {Parameter}
    & {Distribution \& Value}
    & {Description \& Source} \\
  \midrule

  \multicolumn{3}{l}{\textit{Costs in £\,GBP}} \\[2pt]

  Instrument (new)
    & P$(X < 0.1)$: Uniform(200, 300) \newline 
      P$(0.1 \leq X < 0.9)$ = Uniform(600, 700) \newline 
      P$(X > 0.9)$ = Uniform(800, 900) 
    
    & 10\% of the laparoscopic scissors purchased fall in the low price category (between £200 to £300);
    80\% of the laparoscopic scissors lie in the medium price category (between £600 to £700); and
    remaining 10\% in the high cost category ((between £800 to £900). (Source: Industry informed) \\[6pt] 

  Repair
    & Minor repair: £30 \newline
      Major repair: £45
    & (Source: Industry informed) \\[6pt]

  Decontamination (per instrument)
    & £1.05
    & Based on decontamination of a set
      of laparoscopic scissors. (Source: \cite{rizan2022environmental}) \\[6pt]

  \midrule

  \multicolumn{3}{l}{\textit{Emissions in gCO$_2$-eq}} \\[2pt]

  Manufacturing
    & $\mathrm{Triangular}(1013.51,\;1267.4,\;1520.3)$\newline
      $1267.4 \pm 253.4\ (20\%)$\tnote{a}
    & (Mean $= 1267.4$, Source: \cite{Labib2024}) \\[6pt]

  Disposal
    & $\mathrm{Triangular}(277.5,\;346.9,\;416.3)$\newline
      $346.9 \pm 69.38\ (20\%)$\tnote{a}
    & (Mean $= 346.9$, Source: \cite{Labib2024}) \\[6pt]

  Repair (reusable component)
    & $\mathrm{Triangular}(129.12,\;161.4,\;193.68)$\newline
      $161.4 \pm 32.3\ (20\%)$\tnote{a}
    &    (Source: Assumption) \\[6pt]

  \midrule

  \multicolumn{3}{l}{\textit{Miscellaneous}} \\[2pt]

  Laparoscopic scissors weight
    & 323\,g (Plastic: 242\,g;\
      SS\tnote{b}: 82\,g)
    & (Source: \cite{Labib2024}) \\[6pt]

  Order lead time from manufacturer
  to DC
    & $\mathrm{Uniform}(10,\;20)$ days
    & (Source: Industry informed) \\[6pt]

  Probability a laparoscopic scissors requires
  repair, $P(R)$
    & $\mathrm{Uniform}(0.01,\;0.2)$
    & (Source: Industry informed) \\[6pt]

  Probability of major repair given
  laparoscopic scissors failure
    & 0.20
    & (Source: Industry informed) \\[6pt]

  Probability of minor repair given
  laparoscopic scissors failure
    & 0.80
    & (Source: Industry informed) \\[2pt]

  \bottomrule
  \end{tabular}

  \begin{tablenotes}[flushleft]\footnotesize
    \item[a] Modelled via a triangular distribution centred at the mean (mode) with a 20\% variation.
    \item[b] Stainless steel (SS).
  \end{tablenotes}

\end{threeparttable}
\end{table}

\begin{table}[htbp]
\centering
\caption{Single-use instrument-specific data and input parameter
         distributions. \textit{Note: For emission parameters, the first line gives the statistical distribution used in the model (e.g., Triangular(min, mode, max)); the second line gives the source value and the percentage spread applied symmetrically around it to derive the distribution bounds}}
\label{tab:su}
\small
\begin{threeparttable}

  \begin{tabular}{
      p{3.5cm}
      p{4.5 cm}
      p{5.5 cm}
    }

  \toprule
  {Parameter}
    & {Distribution / Value}
    & {Description \& Source} \\
  \midrule

  \multicolumn{3}{l}{\textit{Costs in £\,GBP}} \\[2pt]

  New instrument
    & P$(X < 0.1)$: Uniform(16, 20) \newline 
      P$(0.1 \leq X < 0.9)$ = Uniform(23, 27) \newline 
      P$(X > 0.9)$ = Uniform(27,33)
    & 10\% of the instruments purchased fall in the low price category (between £16 to £20);
    80\% of the instruments lie in the medium price category (between £23 to £27); and
    remaining 10\% in the high cost category ((between £27 to £33). (Source: Industry informed) \\[6pt]

  Remanufactured instrument
    & £12.50 (Fixed) 
    & Available at about half the price of new instruments (Source: Assumption) \\[6pt]

  Average transport distance by
  cargo ship (China to UK)
    & 20{,}921\,km (13{,}000 miles)
    & (Source: Online \cite{sea-distance}) \\[6pt]

  Average transport distance by road
  (port to DC)
    & 100\,km
    & (Source: Assumption) \\[6pt]

  \midrule

  \multicolumn{3}{l}{\textit{Emissions in gCO$_2$-eq}} \\[2pt]

  Manufacturing
    & $\mathrm{Triangular}(528,\;660,\;792)$\newline
      $660 \pm 132\ (20\%)$
    & Modelled via a triangular distribution centered at the mean (mode) with a 20\% variation.
      (Mean = 660, Source: \cite{rizan2022environmental}) \\[6pt]

  Remanufacturing
    & $\mathrm{Triangular}(264,\;330,\;396)$\newline
      $330 \pm 66\ (20\%)$
    & 50\% of manufacturing
      emissions. (Source: Assumption) \\[6pt]

  Disposal
    & $\mathrm{Triangular}(123.2,\;154,\;184.8)$\newline
      $154 \pm 30.8\ (20\%)$
    & Mean $= 154$.
      (Source: \cite{rizan2022environmental}) \\[6pt]

  \midrule

  \multicolumn{3}{l}{\textit{Weight in grams}} \\[2pt]

  Instrument weight
    & 61\,g (Metal: 33.7\,g;
      Plastic: 27.3\,g)
    & (Source: \cite{rizan2022environmental}) \\[6pt]

  \midrule

  \multicolumn{3}{l}{\textit{Miscellaneous}} \\[2pt]

  Order lead time from manufacturer
  to DC
    & $\mathrm{Uniform}(60,\;120)$ days
    & (Source: Industry informed) \\[6pt]

  Order lead time for remanufactured
  instruments
    & $\mathrm{Uniform}(1,\;2)$ days
    & (Source: Assumption) \\[2pt]

  \bottomrule
  \end{tabular}

\end{threeparttable}
\end{table}

\section{Simulation Model Results}
\label{app:add-results}

\subsection{CE Scenario 1: SU Laparoscopic Scissors with Remanufacturing}
\label{app:LE-results}

\begin{table}[htbp]
\centering
\small
\setlength{\tabcolsep}{8pt}
\caption{Tukey pairwise comparison of total supply chain cost (\pounds\,GBP) at
the 5\% significance level for SU laparoscopic scissors with controlled adoption of
remanufacturing (Scenario 1a).}
\label{tab:tukey_1a_cost}
\begin{tabular}{cccccc}
\toprule
{Life cycle} & {1} & {2} & {3} & {4} & {5} \\
\midrule
{1} & ---            & 0               & 0               & 0               & 0               \\
{2} &                & ---             & \textbf{0.220}  & \textbf{0.250}  & 0.020           \\
{3} &                &                 & ---             & \textbf{1.000}  & {0.815}  \\
{4} &                &                 &                 & ---             & \textbf{0.790}  \\
{5} &                &                 &                 &                 & ---             \\
\bottomrule
\end{tabular}
\smallskip
 
\textit{Note:} Means that are not statistically different at the 5\%
level ($p > 0.05$) are shown in \textbf{bold}. A reported value of 0 indicates
$p < 0.001$.
\end{table}

\begin{table}[htbp]
\centering
\small
\setlength{\tabcolsep}{8pt}
\caption{Tukey pairwise comparison of total supply chain emissions
(kg\,CO\textsubscript{2}-eq.) at the 5\% significance level for SU laparoscopic scissors
with controlled adoption of remanufacturing (Scenario 1a).}
\label{tab:tukey_1a_emissions}
\begin{tabular}{cccccc}
\toprule
\textbf{Life cycle} & \textbf{1} & \textbf{2} & \textbf{3} & \textbf{4} & \textbf{5} \\
\midrule
\textbf{1} & --- & 0 & 0 & 0 & 0 \\
\textbf{2} &     & --- & 0 & 0 & 0 \\
\textbf{3} &     &     & --- & 0 & 0 \\
\textbf{4} &     &     &     & --- & 0 \\
\textbf{5} &     &     &     &     & --- \\
\bottomrule
\end{tabular}
\smallskip
 
\textit{Note:} A reported value of 0 indicates $p < 0.001$.
\end{table}

\begin{table}[htbp]
\centering
\small
\setlength{\tabcolsep}{8pt}
\caption{Tukey pairwise comparison of total supply chain cost (\pounds\,GBP) at
the 5\% significance level for SU laparoscopic scissors with prioritised adoption of
remanufacturing (Scenario 1b).}
\label{tab:tukey_1b_cost}
\begin{tabular}{cccccc}
\toprule
\textbf{Life cycle} & \textbf{1} & \textbf{2} & \textbf{3} & \textbf{4} & \textbf{5} \\
\midrule
\textbf{1} & ---            & 0               & 0               & 0               & 0               \\
\textbf{2} &                & ---             & 0               & 0               & 0.020           \\
\textbf{3} &                &                 & ---             & 0               & 0               \\
\textbf{4} &                &                 &                 & ---             & 0               \\
\textbf{5} &                &                 &                 &                 & ---             \\
\bottomrule
\end{tabular}
\smallskip
 
\textit{Note:} A reported value of
0 indicates $p < 0.001$.
\end{table}

\begin{table}[htbp]
\centering
\small
\setlength{\tabcolsep}{8pt}
\caption{Tukey pairwise comparison of total supply chain emissions
(kg\,CO\textsubscript{2}-eq.) at the 5\% significance level for SU laparoscopic scissors
with prioritised adoption of remanufacturing (Scenario 1b).}
\label{tab:tukey_1b_emissions}
\begin{tabular}{cccccc}
\toprule
\textbf{Life cycle} & \textbf{1} & \textbf{2} & \textbf{3} & \textbf{4} & \textbf{5} \\
\midrule
\textbf{1} & --- & 0 & 0 & 0 & 0 \\
\textbf{2} &     & --- & 0 & 0 & 0 \\
\textbf{3} &     &     & --- & 0 & 0 \\
\textbf{4} &     &     &     & --- & 0 \\
\textbf{5} &     &     &     &     & --- \\
\bottomrule
\end{tabular}
\smallskip

 \textit{Note:} A reported value of 0 indicates $p < 0.001$.
\end{table}

\begin{uselscape}

\subsection{CE Scenario 2: Hybrid Laparoscopic Scissors}
\label{app:ce-2-results}

\begin{table}
\centering

\setlength{\tabcolsep}{4pt}
\caption{Tukey pairwise comparison of total supply chain emissions
(kg\,CO\textsubscript{2}-eq.) at the 5\% significance level for hybrid
scissors (Scenario 2).}
\label{tab:tukey_sc2_emissions}
\begin{tabular}{c*{15}{c}}
\toprule
\textbf{Use cycle} &
  \textbf{1} & \textbf{2} & \textbf{3} & \textbf{4} & \textbf{5} &
  \textbf{10} & \textbf{15} & \textbf{20} & \textbf{25} & \textbf{30} &
  \textbf{35} & \textbf{40} & \textbf{45} & \textbf{50} & \textbf{70} \\
\midrule
\textbf{1}  & --- & 0 & 0 & 0 & 0 & 0 & 0 & 0 & 0 & 0 & 0 & 0 & 0 & 0 & 0 \\
\textbf{2}  &  & --- & 0 & 0 & 0 & 0 & 0 & 0 & 0 & 0 & 0 & 0 & 0 & 0 & 0 \\
\textbf{3}  &  &  & --- & 0 & 0 & 0 & 0 & 0 & 0 & 0 & 0 & 0 & 0 & 0 & 0 \\
\textbf{4}  &  &  &  & --- & 0 & 0 & 0 & 0 & 0 & 0 & 0 & 0 & 0 & 0 & 0 \\
\textbf{5}  &  &  &  &  & --- & 0 & 0 & 0 & 0 & 0 & 0 & 0 & 0 & 0 & 0 \\
\textbf{10} &  &  &  &  &  & --- & 0 & 0 & 0 & 0 & 0 & 0 & 0 & 0 & 0 \\
\textbf{15} &  &  &  &  &  &  & --- & 0 & 0 & 0 & 0 & 0 & 0 & 0 & 0 \\
\textbf{20} &  &  &  &  &  &  &  & --- & 0 & 0.034 & 0 & 0 & 0 & 0 & 0 \\
\textbf{25} &  &  &  &  &  &  &  &  & --- & 0 & 0 & 0 & 0 & 0 & 0 \\
\textbf{30} &  &  &  &  &  &  &  &  &  & --- & \textbf{1.000} & \textbf{1.000} & 0 & 0 & 0 \\
\textbf{35} &  &  &  &  &  &  &  &  &  &  & --- & 0 & 0 & 0 & 0 \\
\textbf{40} &  &  &  &  &  &  &  &  &  &  &  & --- & 0 & 0 & 0 \\
\textbf{45} &  &  &  &  &  &  &  &  &  &  &  &  & --- & \textbf{1.000} & \textbf{0.770} \\
\textbf{50} &  &  &  &  &  &  &  &  &  &  &  &  &  & --- & \textbf{0.990} \\
\textbf{70} &  &  &  &  &  &  &  &  &  &  &  &  &  &  & --- \\
\bottomrule
\end{tabular}
\smallskip
 
\textit{Note:} Means that are not statistically different at the 5\%
level ($p > 0.05$) are shown in \textbf{bold}. A reported value of 0 indicates
$p < 0.001$.
\end{table}

\begin{table}
\centering

\setlength{\tabcolsep}{4pt}

\caption{Tukey pairwise comparison of total supply chain cost (\pounds\,GBP) at
the 5\% significance level for hybrid laparoscopic scissors (Scenario 2).}
\label{tab:tukey_sc2_cost}
\begin{tabular}{c*{15}{c}}
\toprule
\textbf{Use cycle} &
  \textbf{1} & \textbf{2} & \textbf{3} & \textbf{4} & \textbf{5} &
  \textbf{10} & \textbf{15} & \textbf{20} & \textbf{25} & \textbf{30} &
  \textbf{35} & \textbf{40} & \textbf{45} & \textbf{50} & \textbf{70} \\
\midrule
\textbf{1}  & --- & 0 & 0 & 0 & 0 & 0 & 0 & 0 & 0 & 0 & 0 & 0 & 0 & 0 & 0 \\
\textbf{2}  &  & --- & 0 & 0 & 0 & 0 & 0 & 0 & 0 & 0 & 0 & 0 & 0 & 0 & 0 \\
\textbf{3}  &  &  & --- & 0 & 0 & 0 & 0 & 0 & 0 & 0 & 0 & 0 & 0 & 0 & 0 \\
\textbf{4}  &  &  &  & --- & 0 & 0 & 0 & 0 & 0 & 0 & 0 & 0 & 0 & 0 & 0 \\
\textbf{5}  &  &  &  &  & --- & 0 & 0 & 0 & 0 & 0 & 0 & 0 & 0 & 0 & 0 \\
\textbf{10} &  &  &  &  &  & --- & 0 & 0 & 0 & 0 & 0 & 0 & 0 & 0 & 0 \\
\textbf{15} &  &  &  &  &  &  & --- & 0 & 0 & 0 & 0 & 0 & 0 & 0 & 0 \\
\textbf{20} &  &  &  &  &  &  &  & --- & 0 & 0.034 & 0 & 0 & 0 & 0 & 0 \\
\textbf{25} &  &  &  &  &  &  &  &  & --- & 0 & 0 & 0 & 0 & 0 & 0 \\
\textbf{30} &  &  &  &  &  &  &  &  &  & --- & 0 & 0 & 0 & 0 & 0 \\
\textbf{35} &  &  &  &  &  &  &  &  &  &  & --- & 0 & 0 & 0 & 0 \\
\textbf{40} &  &  &  &  &  &  &  &  &  &  &  & --- & 0 & 0 & 0 \\
\textbf{45} &  &  &  &  &  &  &  &  &  &  &  &  & --- & 0 & 0 \\
\textbf{50} &  &  &  &  &  &  &  &  &  &  &  &  &  & --- & 0 \\
\textbf{70} &  &  &  &  &  &  &  &  &  &  &  &  &  &  & --- \\
\bottomrule
\end{tabular}
\smallskip
 
\textit{Note:} A reported value of 0 indicates
$p < 0.001$. 
\end{table}

\section{CE Scenario 3: Fully Reusable Laparoscopic Scissors}
\label{app:CE-3-results}

\begin{table}
\centering

\setlength{\tabcolsep}{4pt}
\caption{Tukey pairwise comparison of total supply chain emissions
(kg\,CO\textsubscript{2}-eq.) at the 5\% significance level for fully
reusable laparoscopic scissors (Scenario 3).}
\label{tab:tukey_sc3_emissions}
\begin{tabular}{c*{15}{c}}
\toprule
\textbf{Use cycle} &
  \textbf{1} & \textbf{2} & \textbf{3} & \textbf{4} & \textbf{5} &
  \textbf{10} & \textbf{15} & \textbf{20} & \textbf{25} & \textbf{30} &
  \textbf{35} & \textbf{40} & \textbf{45} & \textbf{50} & \textbf{70} \\
\midrule
\textbf{1}  & --- & 0 & 0 & 0 & 0 & 0 & 0 & 0 & 0 & 0 & 0 & 0 & 0 & 0 & 0 \\
\textbf{2}  &  & --- & 0 & 0 & 0 & 0 & 0 & 0 & 0 & 0 & 0 & 0 & 0 & 0 & 0 \\
\textbf{3}  &  &  & --- & 0 & 0 & 0 & 0 & 0 & 0 & 0 & 0 & 0 & 0 & 0 & 0 \\
\textbf{4}  &  &  &  & --- & 0 & 0 & 0 & 0 & 0 & 0 & 0 & 0 & 0 & 0 & 0 \\
\textbf{5}  &  &  &  &  & --- & 0 & 0 & 0 & 0 & 0 & 0 & 0 & 0 & 0 & 0 \\
\textbf{10} &  &  &  &  &  & --- & 0 & 0 & 0 & 0 & 0 & 0 & 0 & 0 & 0 \\
\textbf{15} &  &  &  &  &  &  & --- & 0 & 0 & 0 & 0 & 0 & 0 & 0 & 0 \\
\textbf{20} &  &  &  &  &  &  &  & --- & 0 & 0 & 0 & 0 & 0 & 0 & 0 \\
\textbf{25} &  &  &  &  &  &  &  &  & --- & \textbf{1.000} & \textbf{0.520} & 0 & 0 & 0 & 0 \\
\textbf{30} &  &  &  &  &  &  &  &  &  & --- & \textbf{0.960} & \textbf{1.000} & 0 & 0 & 0 \\
\textbf{35} &  &  &  &  &  &  &  &  &  &  & --- & 0 & 0 & 0 & 0 \\
\textbf{40} &  &  &  &  &  &  &  &  &  &  &  & --- & \textbf{1.000} & \textbf{1.000} & \textbf{0.950} \\
\textbf{45} &  &  &  &  &  &  &  &  &  &  &  &  & --- & \textbf{1.000} & \textbf{0.990} \\
\textbf{50} &  &  &  &  &  &  &  &  &  &  &  &  &  & --- & \textbf{1.000} \\
\textbf{70} &  &  &  &  &  &  &  &  &  &  &  &  &  &  & --- \\
\bottomrule
\end{tabular}
\smallskip
 
\textit{Note:} Means that are not statistically different at the 5\%
level ($p > 0.05$) are shown in \textbf{bold}. A reported value of 0 indicates
$p < 0.001$. 
\end{table}

\begin{table}
\centering

\setlength{\tabcolsep}{4pt}

\caption{Tukey pairwise comparison of total supply chain cost (\pounds\,GBP) at
the 5\% significance level for fully reusable laparoscopic scissors (Scenario 3).}
\label{tab:tukey_sc3_cost}
\begin{tabular}{c*{15}{c}}
\toprule
\textbf{Use cycle} &
  \textbf{1} & \textbf{2} & \textbf{3} & \textbf{4} & \textbf{5} &
  \textbf{10} & \textbf{15} & \textbf{20} & \textbf{25} & \textbf{30} &
  \textbf{35} & \textbf{40} & \textbf{45} & \textbf{50} & \textbf{70} \\
\midrule
\textbf{1}  & --- & 0 & 0 & 0 & 0 & 0 & 0 & 0 & 0 & 0 & 0 & 0 & 0 & 0 & 0 \\
\textbf{2}  &  & --- & 0 & 0 & 0 & 0 & 0 & 0 & 0 & 0 & 0 & 0 & 0 & 0 & 0 \\
\textbf{3}  &  &  & --- & 0 & 0 & 0 & 0 & 0 & 0 & 0 & 0 & 0 & 0 & 0 & 0 \\
\textbf{4}  &  &  &  & --- & 0 & 0 & 0 & 0 & 0 & 0 & 0 & 0 & 0 & 0 & 0 \\
\textbf{5}  &  &  &  &  & --- & 0 & 0 & 0 & 0 & 0 & 0 & 0 & 0 & 0 & 0 \\
\textbf{10} &  &  &  &  &  & --- & 0 & 0 & 0 & 0 & 0 & 0 & 0 & 0 & 0 \\
\textbf{15} &  &  &  &  &  &  & --- & 0 & 0 & 0 & 0 & 0 & 0 & 0 & 0 \\
\textbf{20} &  &  &  &  &  &  &  & --- & 0 & 0 & 0 & 0 & 0 & 0 & 0 \\
\textbf{25} &  &  &  &  &  &  &  &  & --- & 0 & 0 & 0 & 0 & 0 & 0 \\
\textbf{30} &  &  &  &  &  &  &  &  &  & --- & 0 & 0 & 0 & 0 & 0 \\
\textbf{35} &  &  &  &  &  &  &  &  &  &  & --- & 0 & 0 & 0 & 0 \\
\textbf{40} &  &  &  &  &  &  &  &  &  &  &  & --- & 0 & 0 & 0 \\
\textbf{45} &  &  &  &  &  &  &  &  &  &  &  &  & --- & 0 & 0 \\
\textbf{50} &  &  &  &  &  &  &  &  &  &  &  &  &  & --- & 0 \\
\textbf{70} &  &  &  &  &  &  &  &  &  &  &  &  &  &  & --- \\
\bottomrule
\end{tabular}
\smallskip
 
\textit{Note:} A reported value of 0 indicates
$p < 0.001$. 
\end{table}

 \end{uselscape}
 
\section{Additional Results: Sensitivity Experiments}
\label{app:sensitivity}

\begin{table}[htbp]
\centering
\caption{Full factorial design results for hybrid scissors (Scenario 2):
supply chain cost across all combinations of use cycle, blade cost, and handle
sterilisation cost.}
\label{tab:factorial}
\begin{tabular}{cccc}
\toprule
{Use cycle} &
{Blade cost (\pounds)} &
{Handle sterilisation cost (\pounds)} &
{Supply chain cost (million \pounds)} \\
\midrule
 5 (Low)  & 16 (Low)  & 125 (Low)  &  7.88 \\
70 (High) & 16 (Low)  & 125 (Low)  &  4.65 \\
 5 (Low)  & 34 (High) & 125 (Low)  & 10.68 \\
70 (High) & 34 (High) & 125 (Low)  &  7.45 \\
 5 (Low)  & 16 (Low)  & 400 (High) & 15.77 \\
70 (High) & 16 (Low)  & 400 (High) &  5.32 \\
 5 (Low)  & 34 (High) & 400 (High) & 18.54 \\
70 (High) &34 (High)  & 400 (High) &  8.13 \\
\bottomrule
\end{tabular}
\smallskip
 
\footnotesize\textit{Note:} Full $2^3$ factorial design; costs
represent mean total supply chain cost over the simulation horizon.
\end{table}

\begin{figure}
    \centering
    \includegraphics[width=0.75\linewidth]{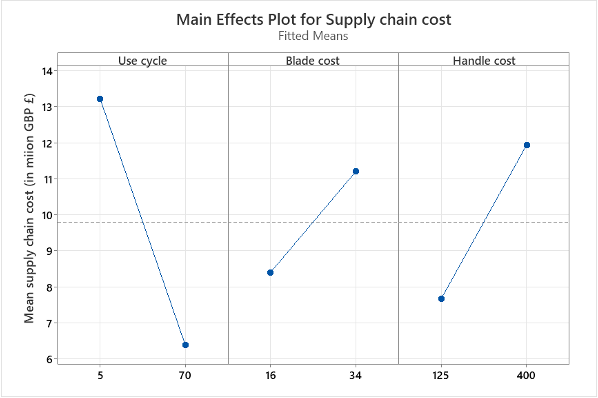}
    \caption{Main effect plots of use cycles, single use blade cost, and reusable handle cost on total supply chain costs (in million \pounds GBP).}
    \label{fig:main-effect}
\end{figure}

\begin{figure}
    \centering
    \includegraphics[width=0.75\linewidth]{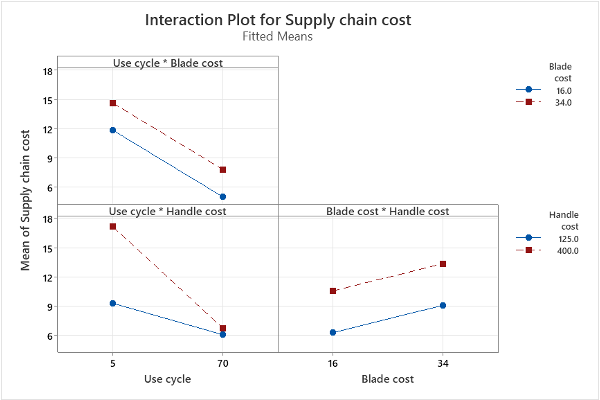}
    \caption{Interaction plots between single use blade costs, reusable handle costs, and use cycles.}
    \label{fig:interaction-effect}
\end{figure}

\end{document}